\documentclass[]{jfm}

\usepackage{epstopdf, epsfig}
\graphicspath{{./pic/}}
\usepackage{color}
\usepackage{amsmath}
\usepackage{bm}% bold math
\usepackage{tabularx}
\usepackage{float}
\usepackage[singlelinecheck=false, width=\linewidth, justification=justified]{caption}
\usepackage{graphicx}
\usepackage{newtxtext}
\usepackage{newtxmath}
\usepackage{natbib}
\usepackage{hyperref}
\usepackage{empheq}
\hypersetup{
	colorlinks = true,
	allcolors = blue,
	citecolor  = black,
}

\newcommand{\RomanNumeralCaps}[1]
\linenumbers

\shorttitle{Inertia and slip effects on the instability of a liquid film coated on a fibre.}
\shortauthor{C. Zhao, R. Qiao, K. Mu, T. Si and X. Luo}

\title{Inertia and slip effects on the instability of a liquid film coated on a fibre}

\author{Chengxi Zhao,
  Ran Qiao, Kai Mu, Ting Si 
 \corresp{\email{tsi@ustc.edu.cn}}
\and Xisheng Luo}

\affiliation{Department of Modern Mechanics, University of Science and Technology of China, Hefei 230026, China}

\begin{document}
\maketitle

\begin{abstract}
To investigate the influence of inertia and slip on the instability of a liquid film on a fibre, a theoretical framework based on the axisymmetric Navier-Stokes equations is proposed via linear instability analysis. 
The model reveals that slip significantly enhances perturbation growth in viscous film flows, whereas it exerts minimal influence on flows dominated by inertia. 
Moreover, under no-slip boundary conditions, the dominant instability mode of thin films remains unaltered by inertia, closely aligning with predictions from a no-slip lubrication model. 
Conversely, when slip is introduced, the dominant wavenumber experiences a noticeable reduction as inertia decreases. 
This trend is captured by an introduced lubrication model with giant slip.
Direct numerical simulations of the Navier-Stokes equations are then performed to further confirm the theoretical findings at the linear stage.
For the nonlinear dynamics, no-slip simulations show complex vortical structures within films, driven by fluid inertia near surfaces.
Additionally, in scenarios with weak inertia, a reduction in the volume of satellite droplets is observed due to slip, following a power-law relationship.
\end{abstract}


%{\bf MSC Codes }  {\it(Optional)} Please enter your MSC Codes here

%% introduction~~~~~~~~~~~~~~~~~~~~~~~~~~~~~~~~~~~~~~~~~~~~~~~~~~~~~~~~~~~~~~~~~~~~~~~~~~~~~~~~~
\section{Introduction}
% try to say film on fibre is important, which has been attracted lots of attentions(chat rewrite )
The investigation of the instability of liquid jets due to surface tension has a long history, tracing back to the pioneering work of \cite{plateau1873} and \cite{rayleigh1878instability,rayleigh1892xvi}. 
This instability phenomenon also holds significant importance in understanding the dynamics of liquid films coated on solid fibres,
with additional complexities at the liquid-solid interface.  
This field has attracted significant scientific attention \citep{quere1999fluid}, owing to its critical relevance across various technological domains, including additive manufacturing \citep{deng2011exploration,oliveira2020revisiting}, droplet transport \citep{lee2022multiple}, chemical element extraction \citep{chen2023spatially} and water collection through fog harvesting \citep{chen2018ultrafast, zhang2022combinational}.

% write the review of the annualr films (1....)
The instability of a film coated on a fibre has been extensively studied across various stages. 
The early-time behaviours can be described by the Rayleigh-Plateau instability \citep{quere1999fluid}, showing that a film with an outer radius $h_0$ becomes unstable to sufficiently long-wavelength disturbances, specifically $\lambda > \lambda_{crit} = 2 \pi h_0$. Here, $\lambda_{crit}$ represents the critical wavelength beyond which the instability ceases to grow. Additionally, the dominant (most unstable/fast growing) modes are influenced by the ratio of $h_0$ to the fibre radius $a$, validated by experiments \citep{goren1964shape}.
Subsequent nonlinear evolution is modelled using the lubrication approximation \citep{hammond1983nonlinear}, resulting in a leading-order lubrication equation that is applicable to films on both the inside and outside of a cylinder.
Due to its simplicity, this lubrication equation and its higher-order versions \citep{craster2006viscous,ruyer2008modelling} have been used in studying various interface dynamics of annular films on fibers. Examples include the transition from absolute unstable regimes to convective ones \citep{kliakhandler2001viscous,duprat2007absolute,craster2009dynamics} and capillary drainage involving complex interactions of `lobes' and `collars' formed on the interfaces \citep{lister2006capillary}.
Recently, these lubrication models have been extended to encompass more complex scenarios by incorporating other physics, such as electric fields \citep{ding2014dynamics}, heat transfer \citep{zeng2017experimental}, thermal fluctuations \citep{zhang2021thermal}, and Van der Waals forces \citep{tomo2022observation}.

% introduce the importance of the slip (2....)
One of the important physical factors is liquid-solid slip, which has recently attracted substantial research interest \citep{secchi2016massive, zhang2020nanoscale, kavokine2021fluids, kavokine2022fluctuation} and been found to influence the dynamics of various interfacial flows \citep{liao2014speeding, halpern2015slip, martinez2020effect, zhao2022fluctuation}.
For the case of cylindrical films, \cite{ding2011stability} introduced a lubrication equation incorporating slip conditions to investigate the instability of films descending along porous vertical fibres. Their findings revealed that the instability is amplified by the presence of a fluid-porous interface, which is modelled using a slip boundary condition.
Regarding annular films within slippery tubes, \cite{liao2013drastic} numerically solved a lubrication equation with leading-order terms, demonstrating that even a fractional amount of wall slip significantly exaggerates the instability, leading to considerably faster drainage compared to the no-slip scenario \citep{hammond1983nonlinear}.
\cite{haefner2015influence} conducted experimental investigations into the influence of slip on the instability for films coated on horizontal fibres. 
Similar to the observations of both \cite{ding2011stability} and \cite{liao2013drastic}, the wall slip was found to enhance the instability, resulting in increased growth rates of perturbations.
The experimental results were also shown to match predictions of a slip-modified lubrication equation.
\cite{halpern2017slip} subsequently illustrated how wall slip can amplify drop formation in a film descending a vertical fiber. 
This observation provides a plausible explanation for the discrepancy between experimentally predicted and theoretically derived critical Bond numbers for drop formation.
More recent investigations delved into the dynamics of films on slippery fibres within non-isothermal conditions \citep{chao2018dynamics} and under the influence of intermolecular forces \citep{ji2019dynamics}, employing more intricate lubrication models.
Despite the extensive use of slip-modified lubrication models, their constraints have been exposed by \citet{zhao2023slip} through linear instability analysis applied to the axisymmetric Stokes equations. The theoretical framework not only highlights an overestimation of the slip-enhanced perturbation growth rate as compared to classical lubrication models, but also reveals a slip-dependent dominant wavelength, deviating from the constant value posited by prior works using the lubrication method \citep{liao2013drastic, haefner2015influence, chao2018dynamics, halpern2017slip}. 

% Then say something about inertia (3....)
% introduce the inertia effect: say something of micro gravity(background)
% try to the cite the paper of inertia of the bounded films
% call back Goren's work, then introduce what we want to do in this work
Noticeably, in modern applications such as additive manufacturing in space \citep{reitz2021additive,van20233d} and 3D printing with liquid metals \citep{assael2010reference,kondic2020liquid},  the role of inertia in governing the dynamics of liquid film interfaces has become notably apparent, contrasting with the predominant neglect of inertia in most preceding investigations.
One exception is the work done by \citet{goren1962instability}, who introduced inertial effects by conducting instability analysis for the full NS equations.
The theoretical findings elucidated distinctions between two limiting cases, i.e. the inviscid case and viscous case without inertia.
However, the impact of inertia within the intermediary regime between these two limits remained uncharted.
\cite{ding2013viscous}, proposed two coupled equations governing the film thickness and flow rate to study instability and dynamics of a film on a fibre, considering both inertia and slip. 
Nonetheless, their focus was primarily on scenarios characterized by small to moderate Reynolds numbers, where the influence of inertia on interface dynamics seemed less pronounced.
While recent studies have extensively investigated the influence of inertia on the instability of planar films \citep{gonzalez2016inertial,moreno2020role}, its effect on the cylindrical films, especially on a fibre with slip, remains unclear, thus motivating the present investigations.

% paper structure
In this work, linear instability analysis of the axisymmetric NS equations is performed to investigate inertia and slip effects on the dynamics of a liquid film on a fibre. Direct numerical simulations of the NS equations are also employed to confirm the theoretical findings and provide more physical insights. 
The article is laid out as follows. 
Non-dimensionalised governing equations for a film on a fibre are introduced in \S\,\ref{sec_MM}. 
Linear instability analysis for the governing equations is performed in \S\,\ref{sec_instability}, where the dispersion relation is derived in \S\,\ref{subsec_dis_relation}, followed by two limiting cases: jet flows in \S\,\ref{subsec_limiting_jet} and film flows without inertia in \S\,\ref{subsec_limiting_stokes} respectively.
Predictions arising from the theoretical model are presented in \S\,\ref{subsec_dis_prediction}.
Subsequently, direct numerical simulations are performed in \S\,\ref{sec_num}. These simulations are compared with the predictions of the theoretical model, specifically concerning the influence on the dominant mode (\S\,\ref{subsec_wavelengths}) and the growth rate (\S\,\ref{subsec_per_growth}) of perturbations.
Nonlinear dynamics extracted from the simulations is also analysed in \S\,\ref{subsec_per_growth}.

% mathematical model ~~~~~~~~~~~~~~~~~~~~~~~~~~~~~~~~~~~~~~~~~~~~~~~~~~~~~~~~~~~~~~~~~~~~~~

\section{Model formulation \label{sec_MM}}

\begin{figure}
\centering
\captionsetup{justification=centering}
\includegraphics[width=0.8\textwidth]{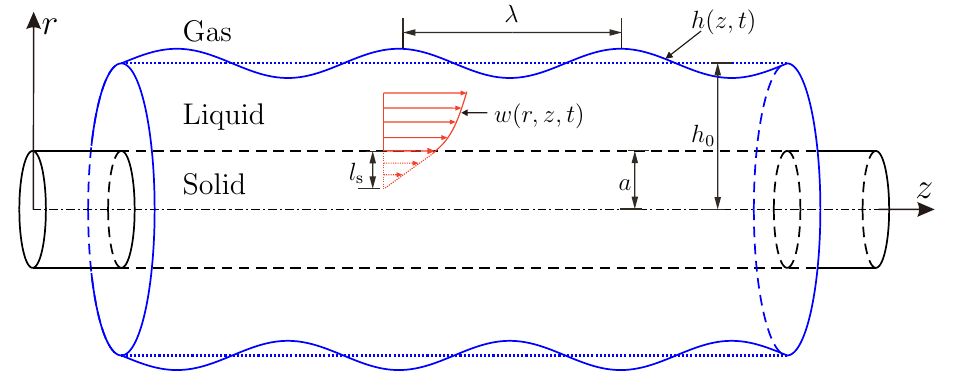}
	\caption{Schematic of a liquid film on a slippery fibre}
\label{fig_schematic}	
\end{figure}

We consider a Newtonian liquid film on a fibre of the radius $a$ with the $z$-axis along the centre line (figure\,\ref{fig_schematic}). 
The initial radius of the film measured from the $z$-axis is $r = h_0$. 
Additionally, gravity is neglected, and we assume uniform external pressure and surface tension.

The incompressible NS equations are employed to predict the dynamics of the flow inside the liquid film.
To identify the governing dimensionless parameters, we non-dimensionalise the NS equations with the rescaling variables shown below:
\begin{equation}
\label{eq_scaling}
(r,z,h)= \frac{(\tilde{r},\tilde{z},\tilde{h})}{h_0}, 
\quad (u,v,w) = \frac{(\tilde{u},\tilde{v},\tilde{w})}{U} , 
\quad t  =\frac{U}{h_0} \tilde{t}, 
\quad p = \frac{h_0}{\gamma}\tilde{p},
\end{equation}
where $\tilde{h}$, $\tilde{t}$, and $\tilde{p}$ represents the dimensional interface height, time and pressure, respectively (note that the dimensional material parameters are not given tildes).
 $(r,\phi,z)$ are the cylindrical coordinates with the corresponding velocities ($u,v,w$).
 $U$ represents the characteristic velocity inside the film and $\gamma$ is the surface tension of the liquid–gas interface.
After eliminating all the derivatives with with respect to $\phi$ and setting $v=0$ in the cylindrical coordinates, the axisymmetric incompressible equations can be written as
\begin{align}
\label{eq_ANS1}
 & \frac{\partial w}{\partial z} + \frac{1}{r} \frac{\partial(ur)}{\partial r}  = 0 \,, \\
   	\label{eq_ANS2}
 & \frac{\partial u}{\partial t} +  u \frac{\partial u}{\partial r} + w \frac{\partial u}{\partial z} 
  = - \frac{1}{We} \frac{\partial p}{\partial r} +
  \frac{1}{Re} \left[ \frac{\partial^2 u}{\partial z^2}+ 
   \frac{\partial}{\partial r} \left(\frac{1}{r} \frac{\partial (ur)}{\partial r}\right) \right]\,,\\
     	\label{eq_ANS3}
   & \frac{\partial w}{\partial t} +  u \frac{\partial w}{\partial r} + w \frac{\partial w}{\partial z} 
  = -\frac{1}{We} \frac{\partial p}{\partial z} + \frac{1}{Re}\left[ \frac{\partial^2 w}{\partial z^2}+ \frac{1}{r}\frac{\partial}{\partial r} \left( r \frac{\partial w}{\partial r} \right)
   \right] \,,
\end{align}
where the non-dimensional quantity $We = \rho U^2 h_0/\gamma$  is the Weber number, which relates
the inertial force to the capillary force.
$Re = \rho U h_0/\mu$ is the Reynolds number, showing the ratio between inertial force and the viscous force.
Here, $\mu$ is the liquid dynamic viscosity, and $\rho$ is the liquid density.

Since the density of gas around the film is much smaller than that of liquid, the gas flow outside can be assumed
to be dynamically passive to simplify the problem. The liquid–gas interface height $h(z, t)$ satisfies the kinematic boundary condition
\begin{equation}
  \label{eq_ANS_kine}
     \frac{\partial h}{\partial t} + w \frac{\partial h}{\partial z} -u = 0 \,.
  \end{equation}
The normal stress balance at the interface $r=h$ gives
\begin{equation}
\label{eq_NS_normal}  p- \frac{We}{Re}\, \bm{n} \cdot \bm{\tau} \cdot
\bm{n} = \nabla \cdot \bm{n} \,,
\end{equation}
where $\tau$ is the shear stress, which is proportional to the strain rate in Newtonian fluids.
$\bm{n}$ is the outward normal and $\nabla \cdot \bm{n}$ represents the dimensionless Laplace pressure.
The tangential force balance is
\begin{equation}
\label{eq_NS_tangential}  \bm{n} \cdot \bm{\tau}  \cdot
\bm{t}=0 \,,
\end{equation}
where $\bm{t}$ is the tangential vector.
With $\bm{n}$ and $\bm{t} $ expressed in terms of the unit vectors in $z$-direction ($\hat{\bm{e}}_z$) and $r$-direction ($\hat{\bm{e}}_r$), 
$$\bm{n} = -\frac{\partial_z h}{\sqrt{1+(\partial_z h)^2}}  \hat{\bm{e}}_z + \frac{1}{\sqrt{1+(\partial_z h)^2}} \hat{\bm{e}}_r \textrm{\,\,\,and\,\,\,} 
\bm{t} = \frac{1}{\sqrt{1+(\partial_z h)^2}}  \hat{\bm{e}}_z + \frac{\partial_z h}{\sqrt{1+(\partial_z h)^2}}  \hat{\bm{e}}_r,$$
%
%With the unit vector in $x$-direction $\bm{n} = (-\partial_x h, 1)/\sqrt{1+\left(\partial_x h \right)^2}$ and unit vector in $r$-direction $\bm{t} = (1, \partial_x h)/\sqrt{1+\left(\partial_x h \right)^2}$,
(\ref{eq_NS_normal}) and  (\ref{eq_NS_tangential}) explicitly give
  \begin{align}
  \label{eq_NS_normal_explicit} 
 &  p - \frac{We}{Re}\frac{2}{1+ \left(\partial_z h \right)^2} 
  \left[  \frac{\partial u}{\partial r} - \frac{\partial h}{\partial z} \left( \frac{\partial w}{\partial r} 
  + \frac{\partial u}{\partial z}\right) + \left(\frac{\partial h}{\partial z} \right)^2 \frac{\partial w}{\partial z} \right]  \nonumber \\
 & =  \frac{1}{h \sqrt{1+\left(\partial_z h \right)^2}}  - \frac{\partial_z^2 h}{\left(1+\left(\partial_z h \right)^2\right) ^\frac{3}{2}} 
  \end{align}
for the normal forces, and
  \begin{equation}
  \label{eq_NS_tangential_explicit}
2 \frac{\partial h}{\partial z} \left( \frac{\partial u}{\partial r} - \frac{\partial w}{\partial z} \right) 
 + \left[ 1- \left( \frac{\partial h}{\partial z} \right)^2 \right]  \left( \frac{\partial w}{\partial r} + \frac{\partial u}{\partial z} \right) = 0 \,.
  \end{equation}
for the tangential forces. Here $\partial_z$ and $\partial_z^2$ refers to the first and second axial derivatives.

In terms of the boundary conditions at the fibre surface $r = \alpha$ ($\alpha$ is the dimensionless fibre radius, i.e. $\alpha = a/h_0$), we introduce the Navier slip boundary condition \citep{navier1823memoire} in tangential direction and the no-penetration boundary condition in the normal direction such that
\begin{align}
\label{eq_slip}
w &= l_s \frac{\partial w}{\partial r}\,, \\
u &= 0 \,.
\label{eq_no_perm}
\end{align}
Here, $l_s$ represents the dimensionless slip length rescaled by $h_0$.

The governing equations above can be further simplified under the long-wave approximation to give lubrication equations. 
Similar to the previous work on planar films \citep{munch2005lubrication}, different rescaling for the leading orders give different forms of the lubrication model.
When inertia is neglected ($Re \ll 1$), the one-equation model, which has been widely used in the previous works \citep{craster2006viscous, haefner2015influence,zhang2020nanoscale,zhao2023slip}, can be obtained from the Stokes equations.
The dimensionless format of the no-slip lubrication model is
\begin{equation}
\label{eq_LE_noslip}
\frac{\partial h}{\partial t} = \frac{1}{ h} \frac{\partial}{\partial z} \left[ M(h) \frac{\partial }{\partial z} 
\left (\frac{1}{h}  - \frac{\partial^2 h}{\partial z^2}  \right)\right]\,,
\end{equation}
where $M(h)$ is the the mobility term
\begin{equation}
M(h) =\frac{1}{16} \left[-3h^4-\alpha^4 + 4 \alpha^2 h^2 + 4 h^4 \mathrm{ln}\left(\frac{h}{\alpha}\right) \right]\,.
\end{equation}
When inertia is not negligible and $l_s \gg 1$, we propose another lubrication model consisting of two equations (see Appendix\,\ref{app_deri_LE} for the derivation).
The dimensionless format of this giant-slip lubrication model is
\begin{subequations}\label{eq_giant_LE}
	\begin{empheq}[left={\empheqlbrace}]{alignat=1}
 \frac{\partial h^2}{\partial t} + \frac{\partial \left(h^2 w \right)}{\partial z} = & 0\,, \\
 \frac{\partial w}{\partial t}+ w \frac{\partial w}{\partial z}  = & -\frac{1}{ We} \frac{\partial}{\partial z}\left(\frac{1}{h}-\frac{\partial^2 h}{\partial z^2} \right)+\frac{1}{ Re}  \frac{3}{h^2-\alpha^2} \frac{\partial \left( h^2 \partial_z w\right)}{\partial z} \notag\\
 &- \frac{1}{ Re} \frac{2 \alpha^2}{h^2-\alpha^2}  \left(\frac{\partial^2 w}{\partial z^2}-\frac{w}{ \alpha\, l_{s}} \right)\,.
	\end{empheq}
\end{subequations}

When $\alpha=0$, the lubrication model for the jet flows \citep{eggers1994drop} is recovered.

\section{Instability analysis \label{sec_instability}}
In this section, linear instability analysis based on equations\,(\ref{eq_ANS1} - \ref{eq_slip}) is performed using the normal mode method, which has been widely used for the instability in different fluid configurations \citep{rayleigh1878instability, tomotika1935instability, craster2006viscous, li2008instability,si2009modes,liang2011linear,gonzalez2016inertial}.

\subsection{Derivation for the dispersion relation \label{subsec_dis_relation}}
To perform instability analysis, the dimensionless perturbed quantities are set as
\begin{align}
u(r,z,t) = \hat{u}(r) e^{\omega t+ikz},\,\, 
w(r,z,t) = \hat{w}(r) e^{\omega t+ikz}  \,\,\textrm{and} \,\,
p(r,z,t) = 1 + \hat{p}(r) e^{\omega t+ikz}\,,
\end{align}
where $\omega$ is the growth rate of perturbations and $k$ is the wavenumber.
Here, we assume that there is no base flow inside the film.
The perturbed quantities are linearly decomposed into the pressure term and the viscosity term,  $\hat{u} =  \hat{u}_p + \hat{u}_{\nu} $ and $\hat{w} = \hat{w}_p + \hat{w}_{\nu} $. 

For the pressure term, the velocity potential $\phi$ (i.e. $\partial_r \phi = \hat{u}_p\textrm{ and } \partial_z \phi = \hat{w}_p$) is introduce to simplify the problem. The mass equation (\ref{eq_ANS1}) becomes a zero-order Bessel equation
\begin{equation}
\label{eq_vel_potential}
\frac{\mathrm{d}}{\mathrm{d} r} \left(r \frac{\mathrm{d} \phi}{\mathrm{d} r} \right) -
k^2 r \phi= 0\,,
\end{equation}
whose solution can be expressed in terms of Bessel functions
\begin{equation}
\label{eq_bessel_phi}
\phi = A_1 \mathrm{I}_0(kr) + B_1 \mathrm{K}_0(kr)\,.
\end{equation}
Here $\mathrm{I}_0$ and $\mathrm{K}_0$ are zero order modified Bessel function of the first and second kinds. 
$A_1$ and $B_1$ are arbitrary constants awaiting determination.
Calculating the derivatives of $\phi$ gives the solution of $\hat{u}_p$ and $\hat{w}_p$
\begin{align}
\hat{u}_p &= k\left[ A_1 \mathrm{I}_1(kr) - B_1 \mathrm{K}_1(kr)\right]\,, \\
\hat{w}_p &= ik\left[ A_1 \mathrm{I}_0(kr) + B_1 \mathrm{K}_0(kr)\right]\,.
\end{align}
Here $\mathrm{I}_1$ and $\mathrm{K}_1$ are first order modified Bessel function of the first and second kinds.
Substituting (\ref{eq_bessel_phi}) into momentum equation (\ref{eq_ANS3}) yields the solution of $\hat{p}$, expressed as
$$\hat{p} = -We\, \omega \left[ A_1 \mathrm{I}_0(kr) + B_1 \mathrm{K}_0(kr)\right] \,.$$

Considering the viscosity term of perturbed quantities, we simplify the momentum equation (\ref{eq_ANS2})  as a first-order Bessel equation,
\begin{equation}
\frac{\mathrm{d}^2 \hat{u}_{\nu}}{\mathrm{d} r^2} + \frac{1}{r}\frac{\mathrm{d} \hat{u}_{\nu}}{\mathrm{d} r} - \left(Re \omega + k^2+\frac{1}{r^2} \right) \hat{u}_{\nu} = 0\,.
\end{equation}
So the solution of $\hat{u}_{\nu}$ is 
\begin{equation}
\hat{u}_\nu =  A_2 \mathrm{I}_1(lr) + B_2 \mathrm{K}_1(lr)\,,
\end{equation}
where $l^2 = k^2+Re\,\omega$. $A_2$ and $B_2$ are another two arbitrary constants.
According to equation\,(\ref{eq_ANS1}),
\begin{equation}
\hat{w}_\nu = \frac{il}{k} \left[ A_2  \mathrm{I}_0(lr) - B_2  \mathrm{K}_0(lr) \right]\,.
\end{equation}

Combining the pressure parts and viscosity parts gives us the general solution of perturbed variables, namely
\begin{empheq}[left={\empheqlbrace}]{alignat=1}
\label{eq_bessel1}
\hat{u} &=  A_1 k \mathrm{I}_1(kr) + A_2 \mathrm{I}_1(lr) - B_1 k \mathrm{K}_1(kr) + B_2 \mathrm{K}_1(lr) \,,\\
\label{eq_bessel2}
\hat{w} &= i \left[ A_1 k \mathrm{I}_0(kr) +  A_2 l \mathrm{I}_0(lr)/k + B_1 k \mathrm{K}_0(kr) - B_2 l \mathrm{K}_0(lr)/k \right]  \,,\\
\label{eq_bessel3}
\hat{p} &=  -We\,\omega \left[ A_1 \mathrm{I}_0(kr) + B_1 \mathrm{K}_0(kr)\right]\,.
\end{empheq}

The dimensionless perturbed quantities for  $\hat{u}$,  $\hat{v}$, $\hat{p}$, combined with $h(x, t) = 1 + \hat{h} e^{\omega t + ikx}$, then are substituted into the boundary equations (\ref{eq_ANS_kine})-(\ref{eq_slip}). For the boundary conditions at the interface ($r = 1$), their linearisation gives
\begin{align}
 \label{eq_NS_BC1}
&  \frac{\mathrm{d} \hat{w} }{\mathrm{d} r }+ ik \hat{u} = 0\,, \\
 \label{eq_NS_BC2}
 &  \hat{p} - 2 \frac{We}{Re}\frac{\mathrm{d} \hat{u}}{ \mathrm{d} r } 
 =  \hat{h}  \left( k^2-1 \right) \,,\\
 \label{eq_NS_BC3}
  & \omega \hat{h} = \hat{u}\,.
\end{align}
And for the boundary conditions on the fibre surface ($r = \alpha$), their linearised forms are
\begin{align}
 \label{eq_NS_BC4}
& \hat{w} = l_{s}\frac{\mathrm{d} \hat{w}}{\mathrm{d} \hat{r}} \,, \\
 \label{eq_NS_BC5}
& \hat{u} = 0 \,.
\end{align}

According to equation\,(\ref{eq_NS_BC3}), $\hat{h}$ in (\ref{eq_NS_BC2}) can be eliminated to give the final four equations of the boundary conditions, i.e. (\ref{eq_NS_BC1}), (\ref{eq_NS_BC2}), (\ref{eq_NS_BC4}) and (\ref{eq_NS_BC5}). 
Substituting the Bessel functions (\ref{eq_bessel1} - \ref{eq_bessel3}) into these perturbed equations leads to a homogeneous system of linear equations for $A_1$, $A_2$, $B_1$ and $B_2$, which has a non-trivial solution only if the determinant of the coefficients vanishes.
In this way, we have the final equation
\begin{equation}
\label{eq_full_dispersion_relation}
\left|
\renewcommand\arraystretch{1.3}
\begin{matrix}
k \mathrm{I}_1(k\alpha) 
& \mathrm{I}_1(l \alpha) 
& - k \mathrm{K}_1(k\alpha) 
& \mathrm{K}_1(l \alpha) 
\\ 
F_{21} & F_{22} & F_{23} & F_{24}
\\
2 k^3 \mathrm{I}_1(k ) 
& (k^2+l^2) \mathrm{I}_1(l) 
& -2 k^3 \mathrm{K}_1(k) 
& (k^2+l^2) \mathrm{K}_1(l) \\
F_{41} & F_{42} & F_{43} & F_{44} \\
\end{matrix} 
\right|=0\,,
\end{equation}
where
\begin{empheq}[left={\empheqlbrace}]{alignat=1}
& F_{21} = k^2 \mathrm{I}_0(k \alpha)-l_{s} k^3 \mathrm{I}_1(k\alpha) \nonumber \,, \\
& F_{22} = l \mathrm{I}_0(l \alpha)-l_{s} l^2 \mathrm{I}_1(l \alpha)  \nonumber \,, \\
& F_{23} = k^2 \mathrm{K}_0(k \alpha)+l_{s} k^3 \mathrm{K}_1(k \alpha) \nonumber \,,\\
& F_{24} = -l \mathrm{K}_0(l \alpha) - l_{s} l^2 \mathrm{K}_1(l \alpha) \nonumber \,, \\
& F_{41} =  \mathrm{I}_0(k) \omega^2 + 2  Re^{-1} k^2 \mathrm{I}'_1(k) \omega + \,We^{-1} (k^2-1) k \mathrm{I}_1(k) \nonumber \,, \\
& F_{42} = 2 Re^{-1} l \mathrm{I}'_1(l) \omega + We^{-1} (k^2-1) \mathrm{I}_1(l) \nonumber \,, \\
& F_{43} =  \mathrm{K}_0(k) \omega^2 - 2 Re^{-1} k^2 \mathrm{K}'_1(k) \omega - We^{-1} (k^2-1) k \mathrm{K}_1(k) \nonumber \,, \\
& F_{44} = 2 Re^{-1} l \mathrm{K}'_1(l) \omega + We^{-1} (k^2-1) \mathrm{K}_1(l) \nonumber\,.
\end{empheq}
As $\omega$ occurs in the argument of  some Bessel functions, such as $\mathrm{I}_1 (l \alpha)$, equation\,(\ref{eq_full_dispersion_relation}) cannot be solved explicitly for $\omega$, except in two limiting cases, which are presented in following subsections.

\subsection{Limiting case of jet flows \label{subsec_limiting_jet}}
When the viscosity of the liquid is neglected ($Re\rightarrow \infty$), the viscosity terms governing the instability become zero, i.e. $u_{\nu}=w_{\nu}=0$. As a result, the dispersion relation is simplified to:
\begin{equation}
\label{eq_dispersion2}
\left|
\renewcommand\arraystretch{1.3}
\begin{matrix}
k \mathrm{I}_1(k \alpha) 
& -k \mathrm{K}_1(k \alpha)  \\
\mathrm{I}_0(k) \omega^2 + \,We^{-1} (k^2-1) k \mathrm{I}_1(k)
&  \mathrm{K}_0(k) \omega^2 - We^{-1} (k^2-1) k \mathrm{K}_1(k) \\
\end{matrix} 
\right|=0\,.
\end{equation}
Here $\omega$ can be expressed explicitly as 
\begin{equation}
\label{eq_inviscid_dispersion_relation}
\omega = \sqrt{\frac{(1-k^2) k}{We}\frac{ \left[ \mathrm{K}_1(k\alpha) \mathrm{I}_1(k) - \mathrm{I}_1(k\alpha) \mathrm{K}_1(k) \right]}{\mathrm{I}_1(k\alpha) \mathrm{K}_0(k)+ \mathrm{K}_1(k\alpha) \mathrm{I}_0(k)}}\,.
\end{equation}
As the fibre radius approaches infinitesimally small values, the flows inside the film are expected to resemble jet flows. Consequently, substituting $\alpha = 0$ into (\ref{eq_inviscid_dispersion_relation}) yields the dispersion relation for the instability of inviscid jets, originally proposed by \cite{rayleigh1878instability}, which is expressed as
\begin{equation}
\label{eq_dis_rayleigh}
\omega = \sqrt{ \frac{(1-k^2) k}{We} \frac{ \mathrm{I}_1(k)}{\mathrm{I}_0(k)}}\,.
\end{equation}

When viscosity is taken into account, relying solely on the condition $\alpha=0$ is no longer adequate to simplify (\ref{eq_full_dispersion_relation})  into the dispersion relation of jet flows. Hence, it becomes imperative to introduce ultra-slip boundary conditions ($l_s \rightarrow \infty$), resulting in
\begin{equation}
\label{eq_dispersion2}
\left|
\renewcommand\arraystretch{1.3}
\begin{matrix}
2 k^3 \mathrm{I}_1(k ) 
& (k^2+l^2) \mathrm{I}_1(l)  \\
F_{41} & F_{42} \\
\end{matrix} 
\right|=0\,.
\end{equation}
This relation can be rearranged as
\begin{equation}
\label{eq_goldin}
\omega^2 + \frac{2}{Re}\frac{k^2}{\mathrm{I}_0(k)}  \left[\mathrm{I}_1'(k) -\frac{2 kl \mathrm{I}_1(k) \mathrm{I}_1'(l)}{(l^2+k^2) \mathrm{I}_1(l)}\right] \omega
= \frac{1}{We} (1- k^2 )k \frac{\mathrm{I}_1(k)}{\mathrm{I}_0(k)} \frac{l^2-k^2}{l^2+k^2}\,,
\end{equation}
which is presented by \cite{goldin1969breakup}.
Using $(l^2+k^2)/(l^2-k^2)  =1+2 k^2/ ( Re \omega)$ and $\mathrm{I}'_1(k) = \mathrm{I}_0(k)-\mathrm{I}_1(k)/k$, we obtain the equivalent representation of (\ref{eq_goldin})
\begin{equation}
\label{eq_lin2}
\omega^2 + \frac{2 k^2}{Re}  \left[2-\frac{\mathrm{I}_1(k)}{k \mathrm{I}_0(k)}+\frac{2 k^2}{l^2-k^2}\left( 1 -\frac{l\, \mathrm{I}_1(k) \mathrm{I}_0(l)}{k\, \mathrm{I}_0(k) \mathrm{I}_1(l)} \right)\right] \omega
= \frac{1}{We} (1- k^2 )k \frac{\mathrm{I}_1(k)}{\mathrm{I}_0(k)} \,,
\end{equation}
which is widely-used form of the dispersion relation for the temporal instability of a viscous Newtonian jet, first proposed by \cite{weber1931zerfall}.
When $Re \rightarrow \infty$, (\ref{eq_dis_rayleigh}) is also recovered.

\begin{figure}
\centering
%\captionsetup{justification=centering}
\includegraphics[width=1.0\textwidth]{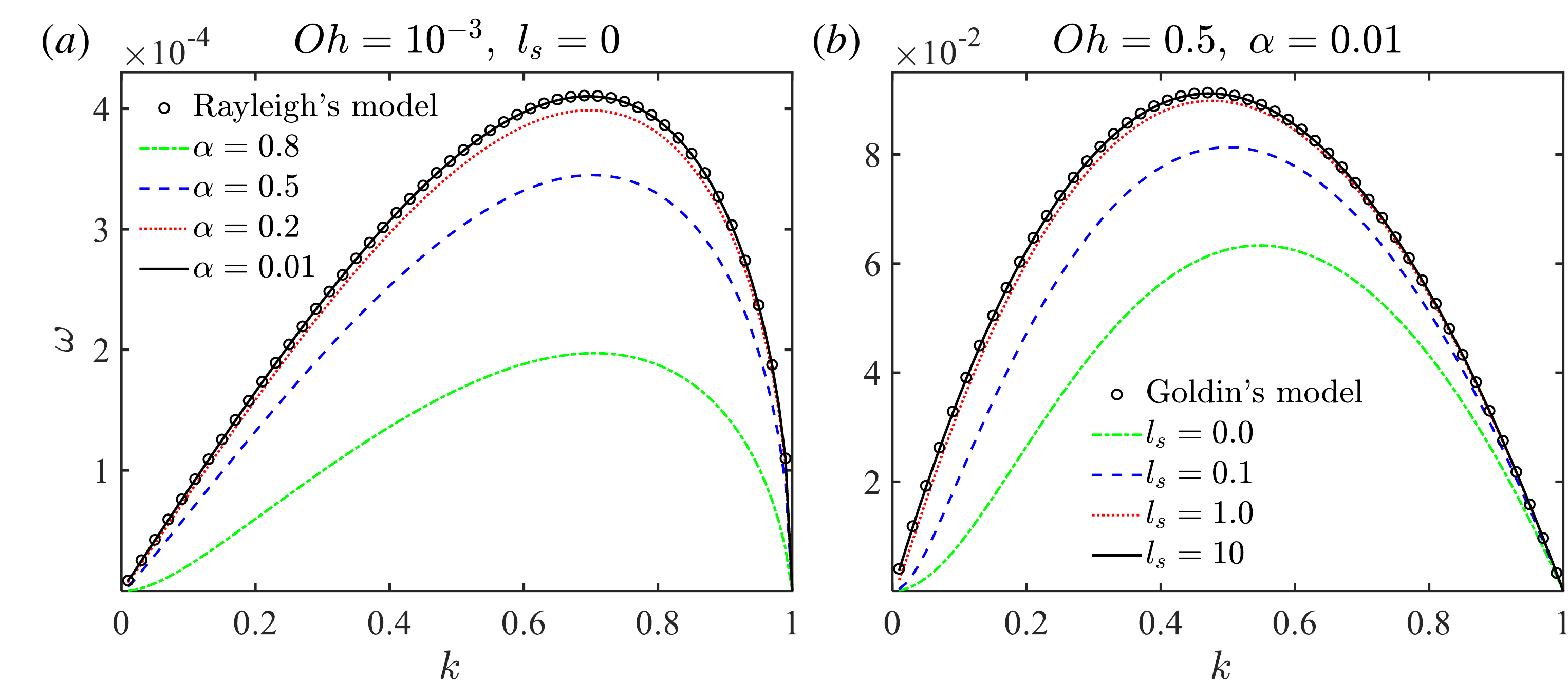}
	\caption{The dispersion relation between the growth rate $\omega$  and the wavenumber $k$ for the limiting cases of inviscid and viscous fluids. (\textit{a}) The inviscid liquid film ($Oh=10^{-3}$) on fibres with different radii  $\alpha=0.8$ (green dash-dotted line), $0.5$ (blue dashed line), $0.2$ (red dotted line), $0.01$  (black solid line); (\textit{b}) The viscous liquid film ($Oh=0.5$) on extremely thin fibres ($\alpha=0.01$) with different slip lengths $l_s=0.0$ (green dash-dotted line), $0.1$ (blue dashed line), $1.0$ (red dotted line), $10$ (black solid line).
	 The circles are the predictions of \cite{rayleigh1878instability} and \cite{goldin1969breakup}. The lines are predictions from the NS dispersion relation (\ref{eq_full_dispersion_relation}). }
\label{fig_dis_jet_limiting}	
\end{figure}

These findings are further confirmed through numerical solutions of equation\,(\ref{eq_full_dispersion_relation}) using the FindRoot function of MATHEMATICA.
In the analysis, the capillary velocity is adopted as the characteristic velocity for non-dimensionalisation, namely $U=\gamma/\mu$. Consequently, we arrive at $Re=We=Oh^{-2}$, where the non-dimensional quantity $Oh = \mu\big/\sqrt{\rho \gamma h_0}$ represents the Ohnesorge number, serving as a linkage between viscous forces, inertial forces, and surface-tension forces.
For inviscid flows, we set $Oh=10^{-3}$. As depicted in figure\,\ref{fig_dis_jet_limiting}\,(\textit{a}), the {results from the NS dispersion relation} (\ref{eq_full_dispersion_relation}) gradually converge towards the predictions of Rayleigh's model as the fibre radius diminishes. This outcome is consistent with the theoretical analysis, thus further validating the numerical solutions of (\ref{eq_full_dispersion_relation}). 
Although the asymptotic behaviours of inviscid cases are realised on the no-slip boundary condition, for the viscous cases, they only manifest when an ultra-slippery fibre is considered, as elucidated in figure\,\ref{fig_dis_jet_limiting}\,(\textit{b}). 
Here, the solutions of (\ref{eq_full_dispersion_relation}) converge towards Goldin's model (\ref{eq_goldin}) as $l_s$ increases. 
This divergence can be attributed to the differential impacts of shear stresses from the fibre surfaces, influenced by the slip length.
% However, these shear stresses have no effects on inviscid flows.
%

%
\subsection{Limiting case of film flows without inertia \label{subsec_limiting_stokes}}

In the regime where inertia of the liquid film is disregarded, i.e. $Re \ll 1$ (or $Oh \gg 1$), $l$ approximates $k$. This leads to the first column in (\ref{eq_full_dispersion_relation}) coinciding with the second column, and the third with the fourth, resulting in an indeterminate form.
To address this issue, we employ the method proposed by \cite{tomotika1935instability}, which involves expanding the Bessel functions in Taylor series with respect to $l$. For instance, $I_1(l) = I_1(k) + I_1'(k) (l-k) + \mathrm{O}\left[(l-k)^2\right]$.
By eliminating the zero-order terms and neglecting higher-order terms (greater than the second order), we arrive at a determinant form.
, expressed as
 \begin{equation}
\left|
\renewcommand\arraystretch{1.3}
\begin{matrix}
k \mathrm{I}_1(k\alpha) & k\alpha \mathrm{I}_1'(k\alpha) & - k \mathrm{K}_1(k\alpha) & k\alpha \mathrm{K}_1'(k\alpha)  
\\
G_{21}
& G_{22}
& G_{23}
& G_{24}
\\
2 k^3 \mathrm{I}_1(k ) 
&2 k^2 \left[ \mathrm{I}_1(k) + k \mathrm{I}_1'(k) 
\right]
& -2 k^3 \mathrm{K}_1(k) 
& 2 k^2 \left[ \mathrm{K}_1(k) + k \mathrm{K}_1'(k) \right]
\\
G_{41} & G_{42} & G_{43} & G_{44} \\
\end{matrix} 
\right|=0.
\label{eq_Stokes_DR1}
\end{equation}
The definitions of the functions $G_{ij}$ can be found in Appendix\,\ref{app_func}. 
Note that $\omega$ appears only linearly in the fourth line of the determinant, the dispersion relation
between $\omega$ and $k$ can be expressed explicitly. After replacing the differentiation of the Bessel functions by $\mathrm{I}'_1(k) = \mathrm{I}_0(k)-\mathrm{I}_1(k)/k$ and $\mathrm{K}'_1(k) = -\mathrm{K}_0(k)-\mathrm{K}_1(k)/k$, we have
\begin{equation}
\label{eq_dispersion_stokes}
\omega = \frac{k^2-1}{2} 
\frac{-\mathrm{I}_1(k) \Delta_1+\mathrm{I}_0(k) \Delta_2 - \mathrm{K}_1(k) \Delta_3 + \mathrm{K}_0(k) \Delta_4  }
{ \left[ k \mathrm{I}_0(k) -\mathrm{I}_1(k) \right] \Delta_1 -k \mathrm{I}_1(k) \Delta_2-\left[ k\mathrm{K}_0(k)+\mathrm{K}_1(k) \right] \Delta_3 + k \mathrm{K}_1(k) \Delta_4   } \,,
\end{equation} 
where details of $\Delta_{ij}$ are shown in Appendix\,\ref{app_func}.
\begin{figure}
\centering
\includegraphics[width=1.0\textwidth]{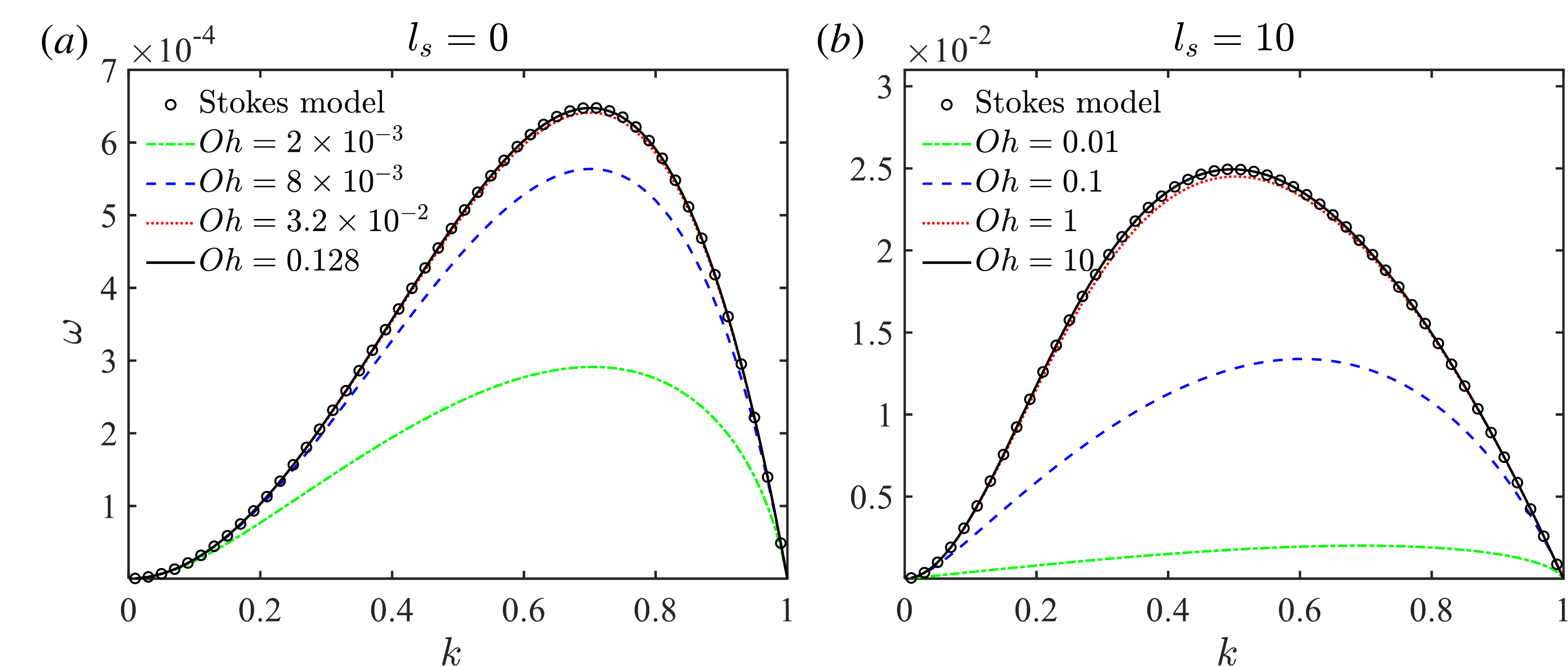}
	\caption{The dispersion relation between the growth rate $\omega$  and the wavenumber $k$ for the limiting cases of thin-film flows ($\alpha = 0.8$). (\textit{a}) No-slip cases with different inertial effects, $Oh=2 \times 10^{-3}$ (green dash-dotted line), $8 \times 10^{-3}$ (blue dashed line), $3.2 \times 10^{-2}$ (red dotted line), $0.128$  (black solid line). (\textit{b}) Slip cases ($l_s=10$) with different inertial effects, $Oh=0.01$ (green dash-dotted line), $0.1$ (blue dashed line), $1$ (red dotted line), $10$  (black solid line).
	 The circles are the predictions from the slip-modified Stokes model \citep{zhao2023slip} and the lines are predictions from the NS dispersion relation (\ref{eq_full_dispersion_relation}).}
\label{fig_limiting_case_stokes}	 
\end{figure}
This dispersion relation (\ref{eq_dispersion_stokes}) is identical to the slip-modified Stokes model proposed by \cite{zhao2023slip}, which was derived directly from the Stokes equations (neglecting inertia in the NS equations).
Numerical investigations are also performed to further support the theoretical analysis, illustrated in figure\,\ref{fig_limiting_case_stokes}, where the predictions generated by (\ref{eq_full_dispersion_relation}) tend to converge towards the Stokes model as $Oh$ increases (inertia declines).
Remarkably, the rate of convergence for the no-slip cases surpasses that of the slip cases significantly. Specifically, when $Oh=3.2 \times 10^{-2}$ (as indicated by the red dotted lines in figure\,\ref{fig_limiting_case_stokes}\,\textit{a}), predictions from the NS dispersion relation closely align with the results of the no-slip Stokes model.
However, for the slip cases, $Oh \geq 1$ is required for a similar convergence.
One plausible explanation for this observation, considering the omission of the base flow, is that the no-slip boundary conditions constrain the fluid motion within the liquid film more effectively than the slip boundary conditions, thereby mitigating the influence of inertia.

\subsection{Predictions of the dispersion relation \label{subsec_dis_prediction}}
Based on the insights gained from our analysis of limiting cases, we turn to the examination of inertia and slip effects in more general scenarios in this subsection.

\begin{figure}
	\centering
	\includegraphics[width=1.0\textwidth]{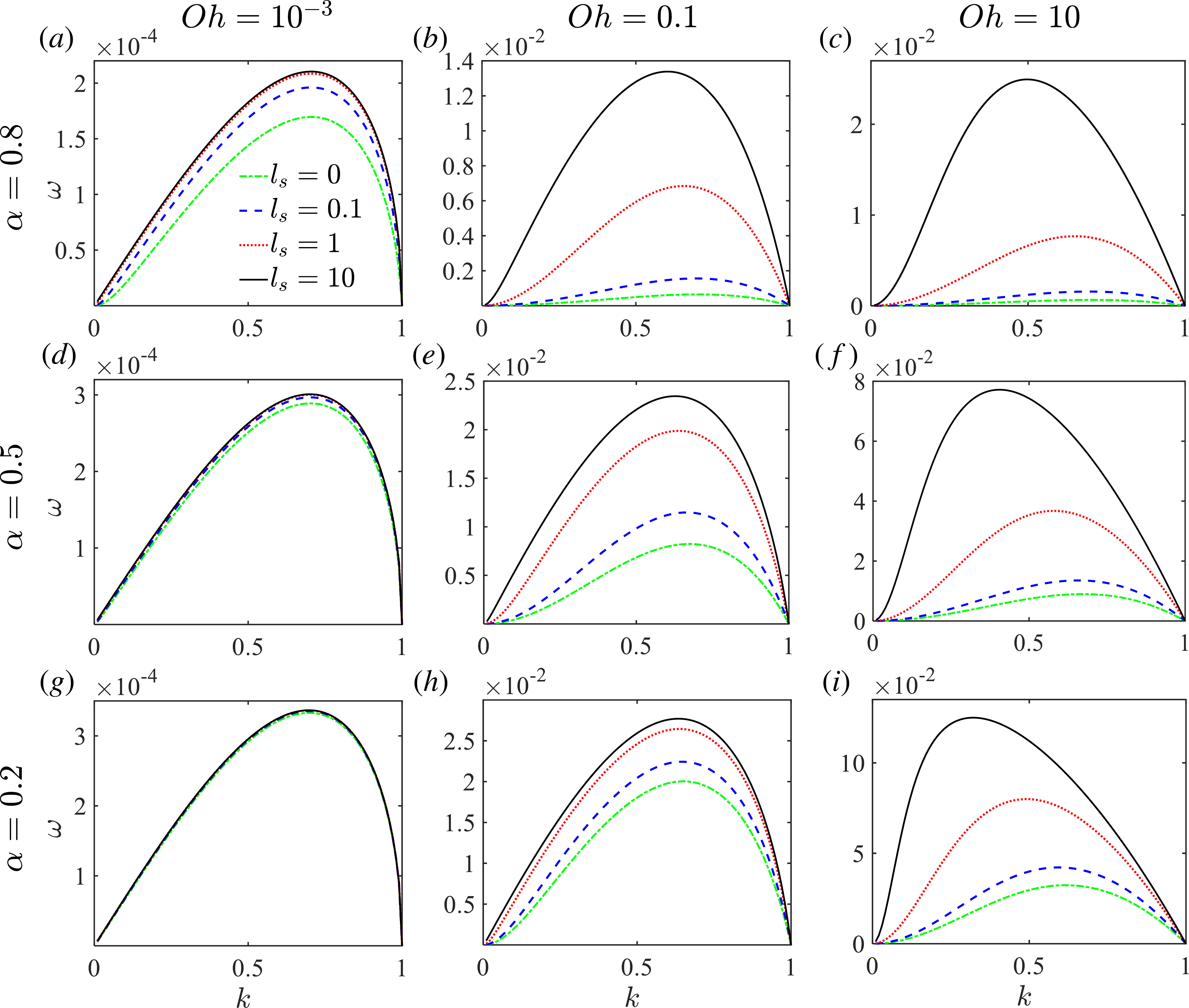}
	\caption{The dispersion relation between the growth rate $\omega$  and the wavenumber $k$ on different boundary conditions of various fibre radii: (\textit{a,b,c}) $\alpha=0.8$, (\textit{d,e,f}) $\alpha=0.5$, (\textit{g,h,i}) $\alpha=0.2$.
		For the inertial effects: (\textit{a,d,g}) $Oh=10^{-3}$, (\textit{b,e,h}) $Oh=0.1$, (\textit{c,f,i}) $Oh=10$.
		Line types represent different values of the slip length: $l_s=0$	(green dash-dotted line), $0.1$ (blue dashed line), $1$ (red dotted line), $10$  (black solid line). }
	\label{fig_dispersion_full_NS}	
\end{figure}

In figure\,\ref{fig_dispersion_full_NS}, we present dispersion relations from (\ref{eq_full_dispersion_relation}) for various slip lengths, while holding specific values of $Oh$ (columns) and $\alpha$ (rows). 
The inertial effects are shown along a given row (fixed $\alpha$), with the first column being inertia-dominated flows and third column corresponding to viscosity-dominated cases. 
The deviations observed across different $l_s$ values indicate that slip predominantly governs the dynamics in viscous flows within the film but has a comparatively minor impact on inertia-dominated cases. This observation is consistent with the outcomes obtained from the limiting case analysis of jet flows (figure\,\ref{fig_dis_jet_limiting}\,\textit{a}).
We also investigate the influence of fibre radii (film thickness) on the instability within each column. As film thickness ($1-\alpha$) increases, the deviations diminish, suggesting that slip effects become less pronounced in thicker films.

These two observations can be explained qualitatively by considering variations in velocity profiles within the liquid films, influenced by both slip and inertia. As the slip length increases, the flow field near the solid wall undergoes a transition from parabolic flow with a non-uniform velocity profile to plug flow with a uniform velocity profile \citep{munch2005lubrication}.
The parabolic flow field decreases and constitutes only a small fraction of the film thickness with an increase in inertia \citep{schlichting1961boundary}, leading to more uniform velocity profiles.
This explains why slip does not significantly impact the instability with $Oh=10^{-3}$ (first column in figure\,\ref{fig_dispersion_full_NS}).
However, when $Oh=10$, most of the flow fields within the films are expected to resemble parabolic profiles, making them more susceptible to the effects of slip.
Moreover, as the film thickness increases, the proportion of the parabolic flow field in the films diminishes, and the velocity profiles become more uniform, resulting in weaker influences of slip on the instability.

% critical wavelengh is not affected, show that capillary  is mot affect
The critical wavenumbers shown in figure\,\ref{fig_dispersion_full_NS} align with the findings of \cite{plateau1873}, i.e. $k_{crit}=2 \pi h_0 / \lambda_{crit}=1$, indicating that slip conditions do not impact these values. 
These values are determined by the interplay between two curvature terms governing the Laplace pressure on the right-hand side of (\ref{eq_NS_normal_explicit}). The circumferential curvature term, $1 \big/ (h \sqrt{1+(\partial_x h )^2} )$, acts as the driving force, while the tangential curvature term, $\partial_x^2 h \big/ (1+\left(\partial_x h \right)^2 ) ^{3/2}$, acts as the resisting force. 
The balance of forces outlined in \eqref{eq_NS_normal_explicit} yields the term $k^2-1$ in $F_{4j}$ of the final dispersion relation \eqref{eq_full_dispersion_relation}, ultimately determining the critical wavenumber as $k_{crit}=1$.

\begin{figure}
\centering
\includegraphics[width=1.\textwidth]{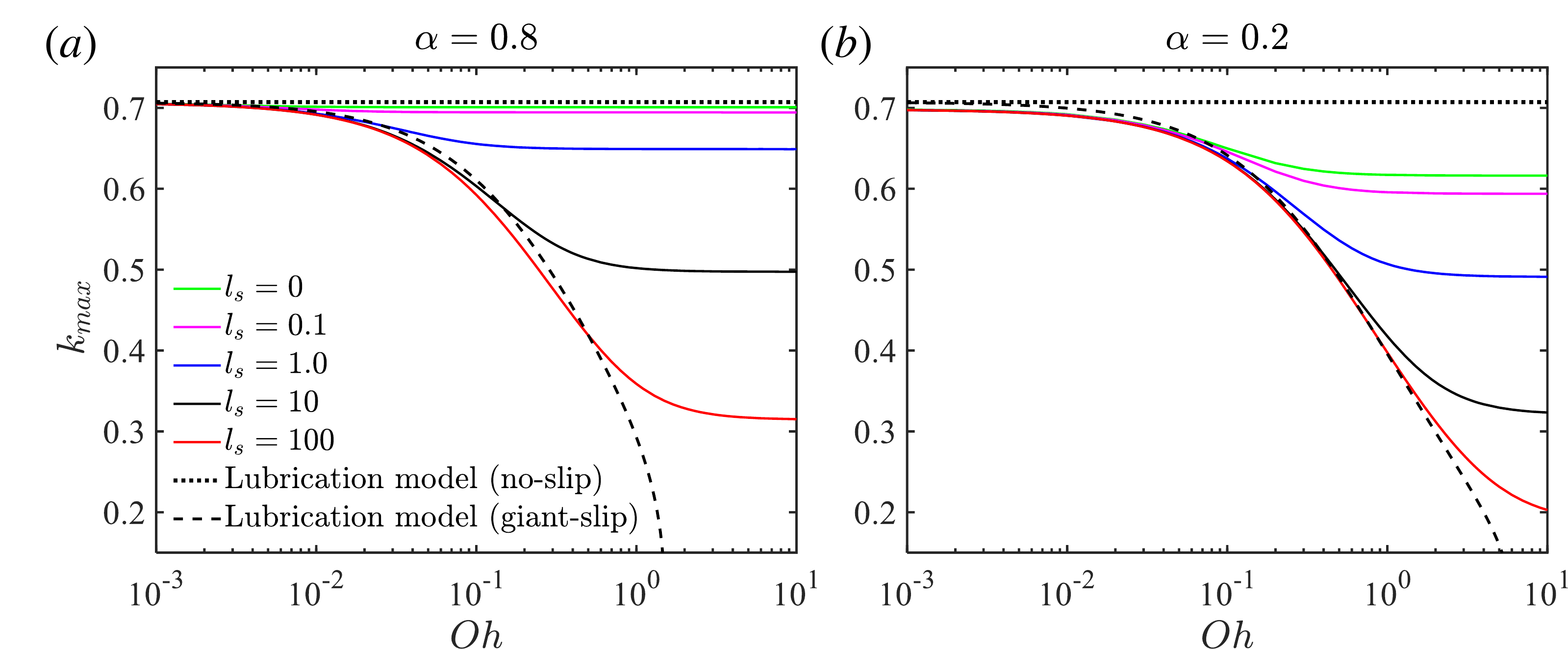}
	\caption{Influence of inertia (different values of $Oh$) on the dominant wavenumber $k_{max}$ on fibres of two radii: (\textit{a}) $\alpha=0.8$, (\textit{b}) $\alpha=0.2$. The solid lines are the predictions of the NS  dispersion relation (\ref{eq_full_dispersion_relation})  for different slip lengths: $l_s=0$ (green), $0.1$ (purple), $1.0$ (blue), $10$ (black), $100$ (red). 
	The dotted lines and dashed lines represent the predictions from the no-slip lubrication model (\ref{eq_LE_noslip}) and the giant-slip one \eqref{eq_giant_LE}}.
\label{fig_kmax}	
\end{figure}

% dominant wavelengths
Figure\,\ref{fig_kmax} further elucidates the relationship between the dominant wavenumber $k_{max}$ and $Oh$. 
Remarkably, inertia appears to exert minimal influence on $k_{max}$ in no-slip cases. Specifically, for a thin film with $\alpha=0.8$ (figure\,\ref{fig_kmax}\,\textit{a}),
$k_{max}$ for small $l_s$ remains unchanged as $Oh$ increases. 
This value is close to the analytical expression $k_{max} = \sqrt{1/2}$ derived from (\ref{eq_dispersion_LE}) for no-slip cases.
This finding offers a plausible explanation for why the no-slip lubrication model (\ref{eq_LE_noslip}), which neglects inertia, has demonstrated remarkable capabilities in predicting the wavelengths observed in numerous experimental studies \citep{quere1990thin,duprat2007absolute,craster2009dynamics, craster2006viscous, ji2019dynamics}.
Conversely, in slip cases, $k_{max}$ exhibits a decline with increasing $Oh$. This trend holds across different values of slip length ($l_s$), with more pronounced decreases in $k_{max}$ for larger slip lengths. Encouragingly, the predictions of the giant-slip lubrication model \eqref{eq_dispersion_giant_slip} closely align with the results of cases with substantial slip ($l_s>10$).
As $Oh$ becomes sufficiently large for all $l_s$, $k_{max}$ predicted by the NS dispersion relation converges to a constant, whereas in the giant-slip lubrication model, $k_{max}$ consistently decreases rapidly.
For a thick film with $\alpha=0.2$ (figure\,\ref{fig_kmax}\,\textit{b}), though $k_{max}$ for the no-slip case decline from $0.7$ to $0.61$, which cannot be predicted by the no-slip lubrication model \eqref{eq_dispersion_LE}, the variation trend of $k_{max}$ with $Oh$ is similar to that observed in thin films ($\alpha=0.8$).
Furthermore, slip is found not to significantly impact $k_{max}$ in film flows dominated by inertia ($Oh<10^{-2}$), consistent with the findings in figure\,\ref{fig_dispersion_full_NS}. However, in viscous cases ($Oh>1$), $k_{max}$ decreases significantly as $l_s$ increases, corroborating the conclusions drawn in the work of \cite{zhao2023slip}.

\section{Direct numerical simulations \label{sec_num}}
In this section, direct numerical simulations are performed to corroborate  the theoretical findings in \S\,\ref{sec_instability} and gain deeper physical understanding of how inertia and slip impact the instability of films on fibres.

The numerical solution of the NS equations is achieved through the Finite Element method within a computational framework facilitated by COMSOL Multiphysics 6.1. 
The simulations for the film flows are conducted using the Arbitrary Lagrangian–Eulerian (ALE) approach. In this method, free surface nodes are moved in a Lagrangian manner, deforming the computational domain, while nodes inside the film follow a predefined evolution.  
This approach offers a distinct advantage over other techniques, such as level set or phase field methods. Unlike these alternatives, the ALE approach ensures an exact capture of the free surface, akin to the Lagrangian approach, while retaining the primary advantage of Eulerian methods that mesh elements are less prone to distortion. Consequently, it has gained widespread use in predicting the dynamics of free-surface flows across various phenomena such as droplet dynamics \citep{chubynsky2020bouncing, chakraborty2022computational}, dynamics of a ligament \citep{wei2021statics}, jet breakup \citep{martinez2020natural} and the instability of planar films \citep{gonzalez2016inertial,moreno2020stokes}.
However, it is worth noting that this method is not suitable for scenarios where the topology of the domain might change, such as in cases involving fluids inside the film after rupture. This limitation arises from the necessity to maintain consistent mesh connectivity throughout the simulation, and the only workaround is to manually change the mesh topology.

\begin{figure}
\centering
\includegraphics[width=1.0\textwidth]{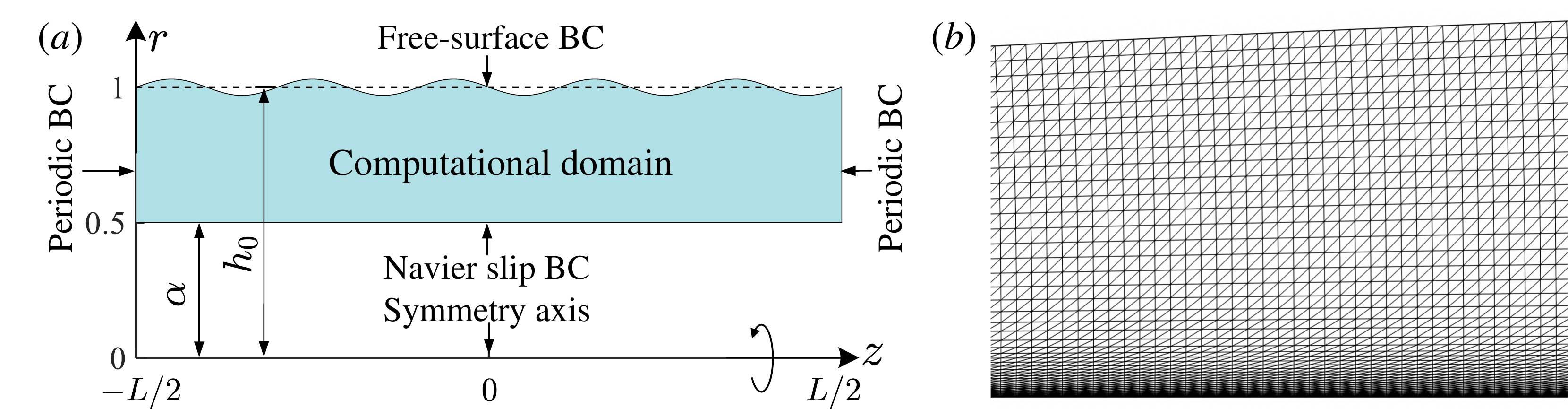}
	\caption{Numerical settings: (\textit{a}) quadrilateral computational domains with different boundary conditions (BCs); (\textit{b}) non-uniform triangular mesh.}
\label{fig_numerics_setting}	
\end{figure}

The computational domain is a quadrilateral (the section of a hollow fibre in cylindrical coordinates) with a size $[\alpha, h_0+\hat{h}] \times [0, L]$, illustrated in figure\,\ref{fig_numerics_setting}\,(\textit{a}). 
Here $L$ is the length of the film/fibre.  
we assign a value of $\alpha=0.5$ to the radius of the fibre, and the initial radius of the film is $h_0 = 1$. Small perturbations ($\hat{h}$) are introduced at the liquid-gas interface.
The left and right boundary conditions of the computational  domain are considered periodic. The top represents the free surface, following \eqref{eq_ANS_kine}-\eqref{eq_NS_tangential}.
The bottom is treated as the slip-wall boundary modelled by \eqref{eq_slip} and \eqref{eq_no_perm}, where the slip length $l_s$ serves as an input parameter for this boundary condition.
The axial velocity on the boundary $w_b = l_s \partial_r w_b$. 
When $l_s=0$, the no-slip boundary condition ($w_b=0$) is recovered. 
The computational mesh for the liquid domain utilises non-uniform triangular Lagrange elements, shown in figure\,\ref{fig_numerics_setting}\,(\textit{b}). Special attention is given to placing finer grid elements near the solid boundary to accurately capture the fluid behaviours in the non-uniform velocity profiles.
The minimum grid size employed is $10^{-3}$.
Additionally, the variable-order backward differentiation formula is utilised for the temporal integration of the NS equations.
All simulations are conducted using dimensionless units, which are established through rescaling variables as described in equation\,(\ref{eq_scaling}).

We explore two different configurations in this study, each varying in film length and initial perturbations, allowing us to investigate the combined effects of inertia and slip on the dominant wavelength (\S\,\ref{subsec_wavelengths}) and the evolution of perturbation growth (\S\,\ref{subsec_per_growth})

\subsection{Dominant wavelengths of perturbations \label{subsec_wavelengths}}

To investigate the influence of inertia and slip effects on the dominant wavelengths of perturbations, we perform simulations involving long films with a length of $L=200$ on various slippery fibres. We explore twenty different cases by considering five values of the Ohnesorge number, specifically $Oh=10^{-3}$, $10^{-2}$, $0.1$, $1$, and $10$, on fibres characterized by four slip lengths, namely $l_\mathrm{s}=0$, $1$, $10$, and $100$.
In these simulations, the system is initiated with random initial perturbations described as $h(z,0) = 1+ \varepsilon N(z)$, where $\varepsilon = 10^{-3}$, and $N(z)$ is a random variable following a normal distribution with a mean of zero and a unit variance.
These initial perturbations are designed to replicate the arbitrary disturbances commonly encountered in reality.

\begin{figure}
\centering
\includegraphics[width=0.85\textwidth]{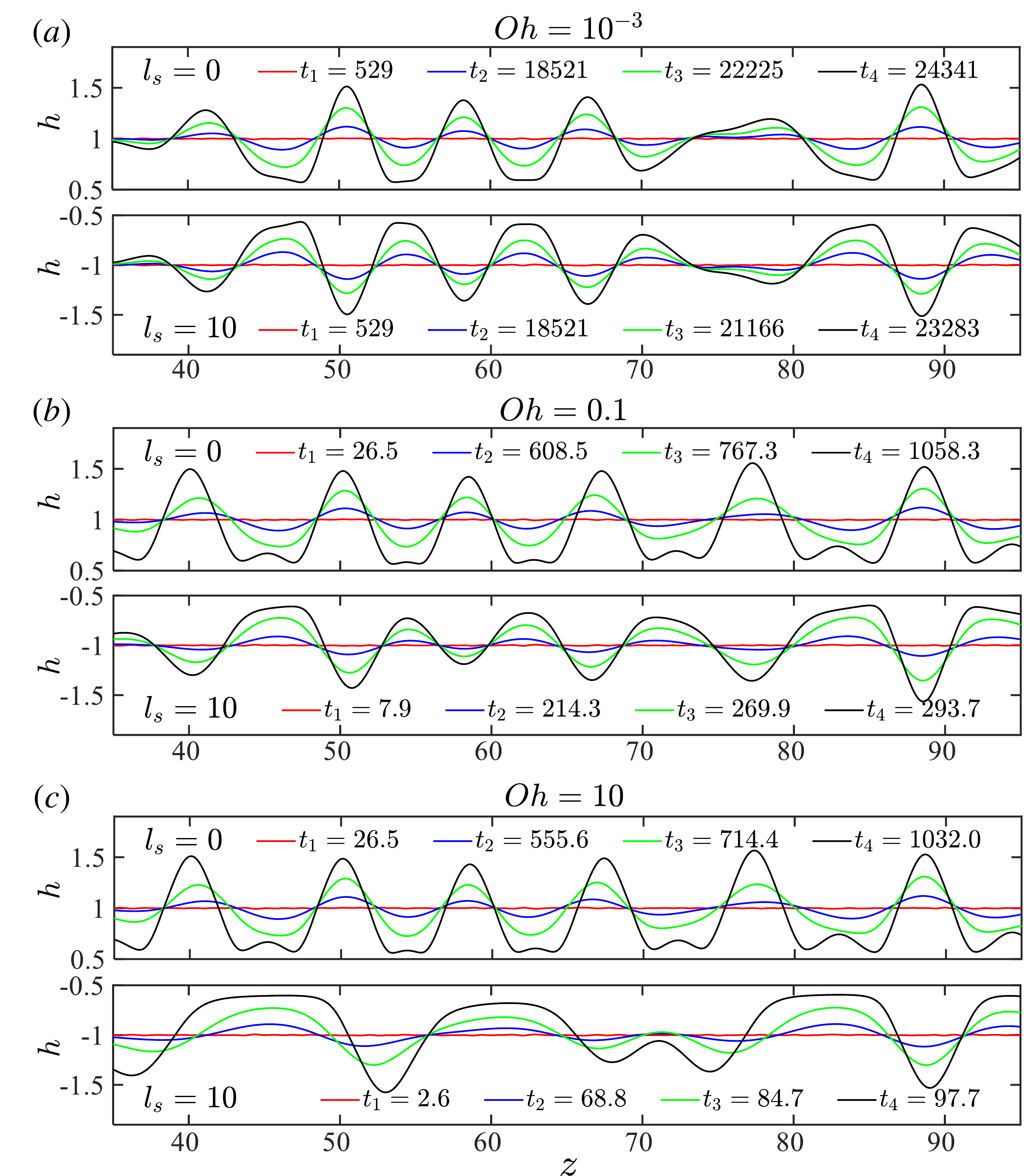}
	\caption{ Interface profiles at four time instants, illustrated in different colours, on the fibres of the radius $\alpha=0.5$.
	The inertial effects are presented by different values of $Oh$: (\textit{a}) $Oh=10^{-3}$, (\textit{b}) $Oh=0.1$, (\textit{c}) $Oh=10$. 
	The upper panels in each figure are the predictions of no-slip cases ($l_s=0$) and lower ones are the results of slip cases ($l_s=10$).}
\label{fig_wave_display}	
\end{figure}

Driven by surface tension, small random perturbations gradually evolve over time, giving rise to significant capillary waves, as illustrated in figure\,\ref{fig_wave_display}, which shows the evolution of interface profiles $h(z,t)$ for six different cases. To assess the impact of inertial effects, identical initial conditions are assigned for all six cases.
For the inertia-dominated cases ($Oh=10^{-3}$), the evolution of capillary waves in slip cases is nearly indistinguishable from that in no-slip cases, except for the slightly faster growth of perturbations in the slip case compared to the no-slip case (figure\,\ref{fig_wave_display}\,\textit{a}).
Conversely, when viscosity becomes significant (figure\,\ref{fig_wave_display}\,\textit{b,c}), slip noticeably affects $h(z,t)$, resulting in faster perturbation growth and longer capillary waves.
Furthermore, the wavelengths of the no-slip cases do not appear to be significantly influenced by inertia (see the six waves in the upper panels of figure\,\ref{fig_wave_display}\,\textit{a,b,c}), despite the presence of deviations in local interface profiles.
All these findings align qualitatively with the theoretical predictions outlined in \S\,\ref{subsec_dis_prediction}.
Additionally, we conduct a comparison between the interface profiles obtained through simulations for the NS equations and those calculated numerically from the lubrication models. The details are presented in Appendix\,\ref{app_com_int}.

\begin{figure}
\centering
\includegraphics[width=1.0\textwidth]{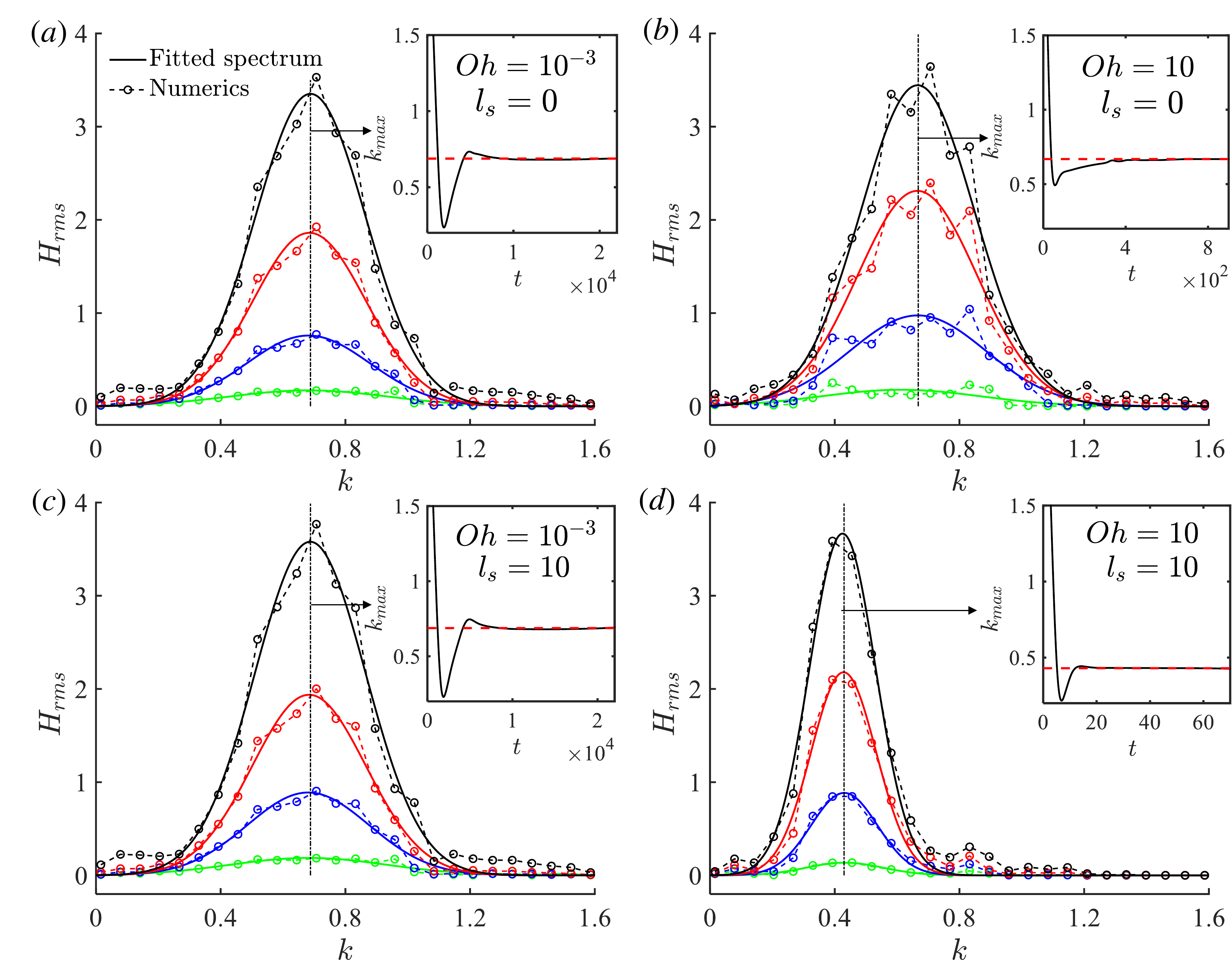}
	\caption{The root mean square (rms) of nondimensional perturbation amplitude versus nondimensional
wave number on fibres with two slip lengths at four time instants: (\textit{a}) $10583$ (green), $15875$ (blue), $19050$ (red) and $21166$ (black); 
(\textit{b}) $264.6$ (green), $476.2$ (blue), $582.1$ (red) and $635.0$ (black);
(\textit{c})  $10583$ (green), $15875$ (blue), $18521$ (red) and $20632$ (black);
(\textit{d}) $26.5$ (green), $52.9$ (blue), $66.1$ (red) and $74.1$ (black). 
The circles are extracted from numerical simulations fitted by the Gaussian function.
The inset shows the time history of the dominant wavenumbers extracted from the spectra.}
\label{fig_spectrum}	
\end{figure}

To quantitatively compare the numerical observations with the theoretical predictions derived from (\ref{eq_full_dispersion_relation}), we conduct multiple independent simulations (10 for each case) with different initial conditions to collect statistical data of the dominant modes. This statistical approach was proposed by \cite{zhao2019revisiting} and has been employed in evaluating dominant modes of instability in various films \citep{zhao2021influence,zhao2023fluctuation,zhao2023slip}.
For each simulation, a discrete Fourier transform is applied to the interface position $h(z,t)$ to obtain the power spectral density (PSD) of the perturbations. The square root of the ensemble-averaged PSD ($H_{rms}$) at each time step is depicted in figure\,\ref{fig_spectrum}, with a Gaussian function used to fit the modal distribution (spectrum). The peak of this fitted spectrum corresponds to the dominant wavenumber $k_{max}$, as indicated by the black dash-dotted lines.
Extracting $k_{max}$ from the fitted spectrum at each time instant yields the insets in figure\,\ref{fig_spectrum}.
Promisingly, $k_{max}$ converges to a constant rapidly in all the cases.
Consistent with the findings in figure\,\ref{fig_wave_display}, the two spectra in figure\,\ref{fig_spectrum}\,(a,c) are nearly identical, suggesting that slip does not significantly impact the instability in the inertia-dominated regime. However, in figure\,\ref{fig_spectrum}\,(\textit{d}), the spectrum for the slip case exhibits a smaller dominant wavenumber compared to the no-slip case in figure\,\ref{fig_spectrum}\,(\textit{c}). This discrepancy indicates that, in the viscous regime, slip leads to an increase in the wavelength.

\begin{figure}
\centering
\includegraphics[width=0.7\textwidth]{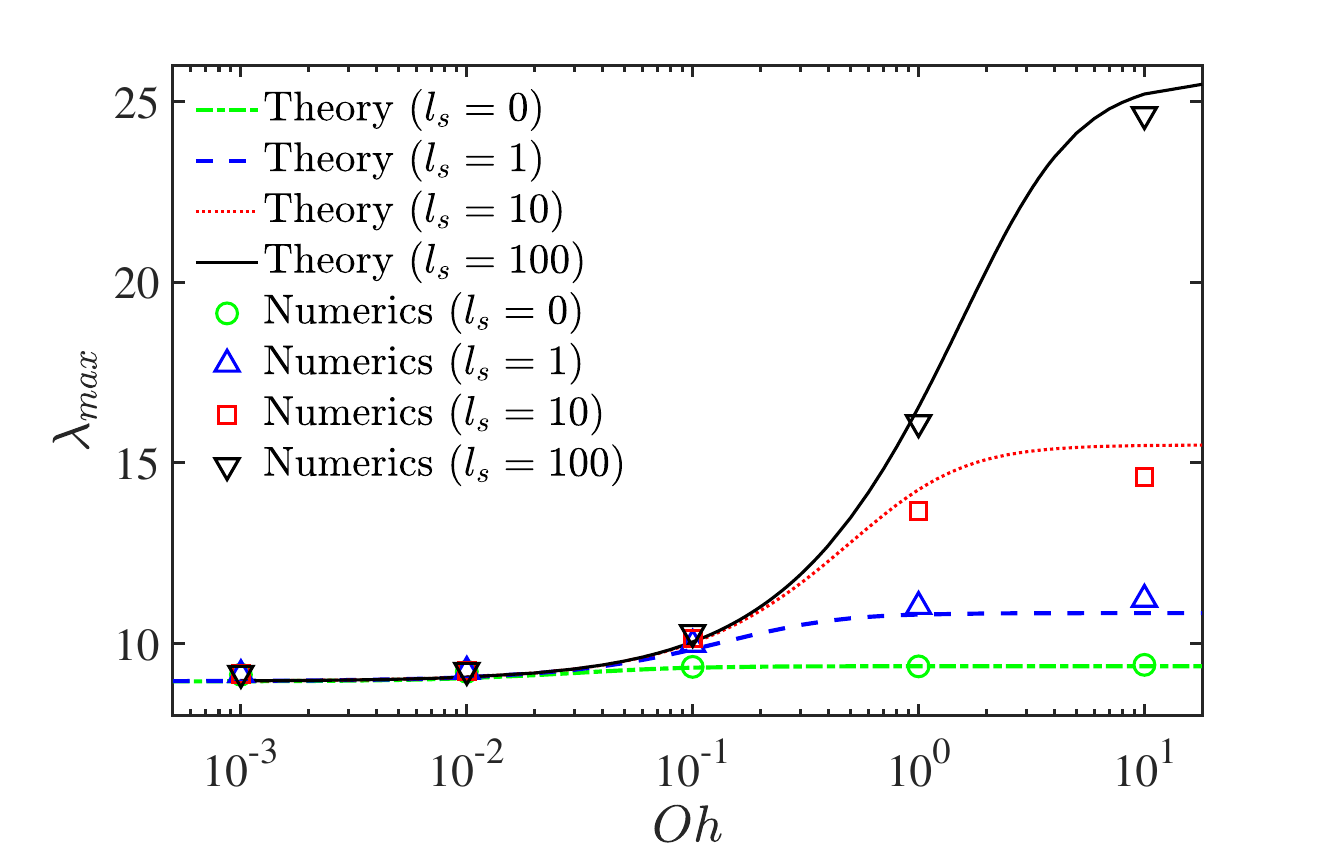}
	\caption{Inertial effects on the dominant wavelengths: a comparison between the theoretical predictions of (\ref{eq_full_dispersion_relation}) and numerical solutions with different slip lengths.
}
\label{fig_dominant_wavelengths}	
\end{figure}

This statistical analysis is applied for all the cases to generate the symbols ($\lambda_{max} = 2 \pi/k_{max}$) in figure\,\ref{fig_dominant_wavelengths}, which exhibit a good agreement with theoretical predictions.
Consequently, we can draw the conclusion that the dominant modes of the thin-film instability remain largely unaffected by inertia in no-slip cases. However, they become significantly influenced by inertia on slippery surfaces, leading to the formation of longer perturbation waves.

\subsection{Evolutions of perturbation growth \label{subsec_per_growth}}
To investigate the impact of inertia and slip on perturbation growth, we conduct simulations of a relatively short film with $L=10$ on slippery fibres. The film is initially perturbed with $h(z,t) =1+ \varepsilon \mathrm{cos}\left[2 \pi (z/L-1/2)\right]$, where $\varepsilon=0.01$.

\begin{figure}
\centering
\includegraphics[width=1.0\textwidth]{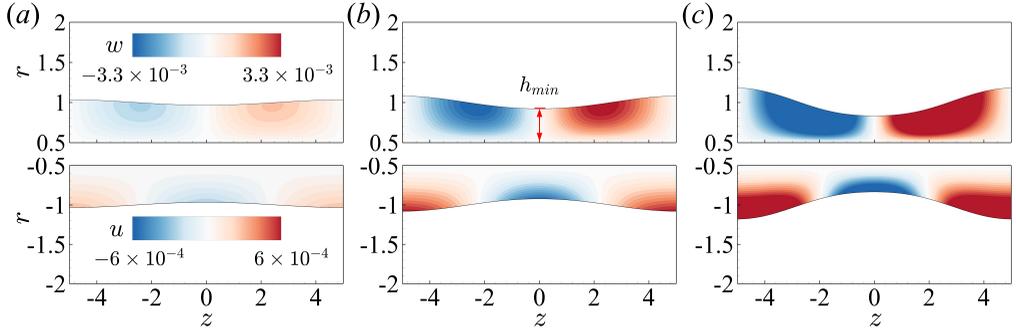}
	\caption{Film thinning of one perturbation wave in an axially symmetric domain. 
For this case, $L=10$, $\alpha=0.5$, $Oh=0.1$ and $l_s=0$.
	Contours of the axial velocity $w(r,z,t)$ (upper half) and radial velocity $u(r,z,t)$ (lower half)  inside the film are shown at three time instants: (\textit{a}) $t_1=158$; (\textit{b}) $t_2=264$; (\textit{c}) $t_3=370$. $h_\mathrm{min}$ represents the minimum radius of the film.}
\label{fig_flow_filed_linear}	
\end{figure}

Figure\,\ref{fig_flow_filed_linear} presents the time evolution of the film interface on a no-slip fibre with a radius of $\alpha=0.5$, with the corresponding fluid structure.
The contour in the lower panel of figure\,\ref{fig_flow_filed_linear} depicts the radial velocity $u$. 
The upper half displays the axial velocity $w$, showing that opposite fluxes occur, directed towards the left and right boundaries, as the perturbation at the free surface grows due to instability.
Notably, while the magnitudes of both $u$ and $w$ increase during the process, the fluid structure remains similar at the linear stage (the amplitude of disturbances is typically less than 20\% of initial radius, i.e. $\hat{h}<0.2$), as evident from the contour distribution in figure\,\ref{fig_flow_filed_linear}.

\begin{figure}
\centering
\includegraphics[width=1.0\textwidth]{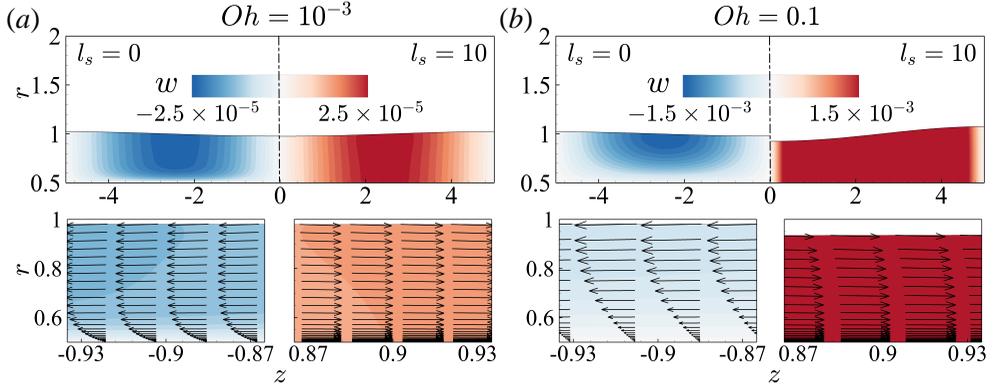}
	\caption{Inertia and slip effects on axial velocity fields: (\textit{a}) $Oh=10^{-3}$ at $t=2027$, (\textit{b}) $Oh=0.1$ at $t=106$. Here, the fibre radius $\alpha = 0.5$.
	The left panels in each figure illustrate the contours of the entire configuration (upper half) and velocity vectors of the local field near $\left|z \right|=0.9$ (lower half) of the no-slip cases.
	The right panels show the results of the slip case ($l_s=10$).
}
\label{fig_velocity_vector}	
\end{figure}

In figure\,\ref{fig_velocity_vector}, we present more velocity fields for four distinct cases, considering two different values of $Oh$: $10^{-3}$ for the inertia-dominated cases and $0.1$ for the viscous cases, on both no-slip (left panel) and slip (right panel) fibres. 
According to the variations of the contours in the upper half of figure\,\ref{fig_velocity_vector}\,(\textit{b}), slip not only alters the velocity distribution near the surface of the fibres but also accelerates instability in the viscous cases, as evidenced by the larger axial velocity component $w$. However, in the case of inertia-dominated flow, slip has a negligible impact on instability, corroborating our previous findings.
% introduce the velocity vector
The lower half of figure\,\ref{fig_velocity_vector} illustrates the velocity vectors near $z=\pm0.9$ for different cases. 
It is apparent that slip reduces the velocity gradient $\partial_r w$ near the surface. 
Furthermore, the parabolic flow field in the no-slip case with $Oh=10^{-3}$ is observed to be considerably smaller than that in the case with $Oh=0.1$. This finding lends additional support to the explanation provided following figure\,\ref{fig_dispersion_full_NS}. In essence, the more uniform velocity profile in the inertia-dominated case serves to limit the impact of slip, resulting in nearly identical dynamics of the instability.

\begin{figure}
\centering
\includegraphics[width=1.0\textwidth]{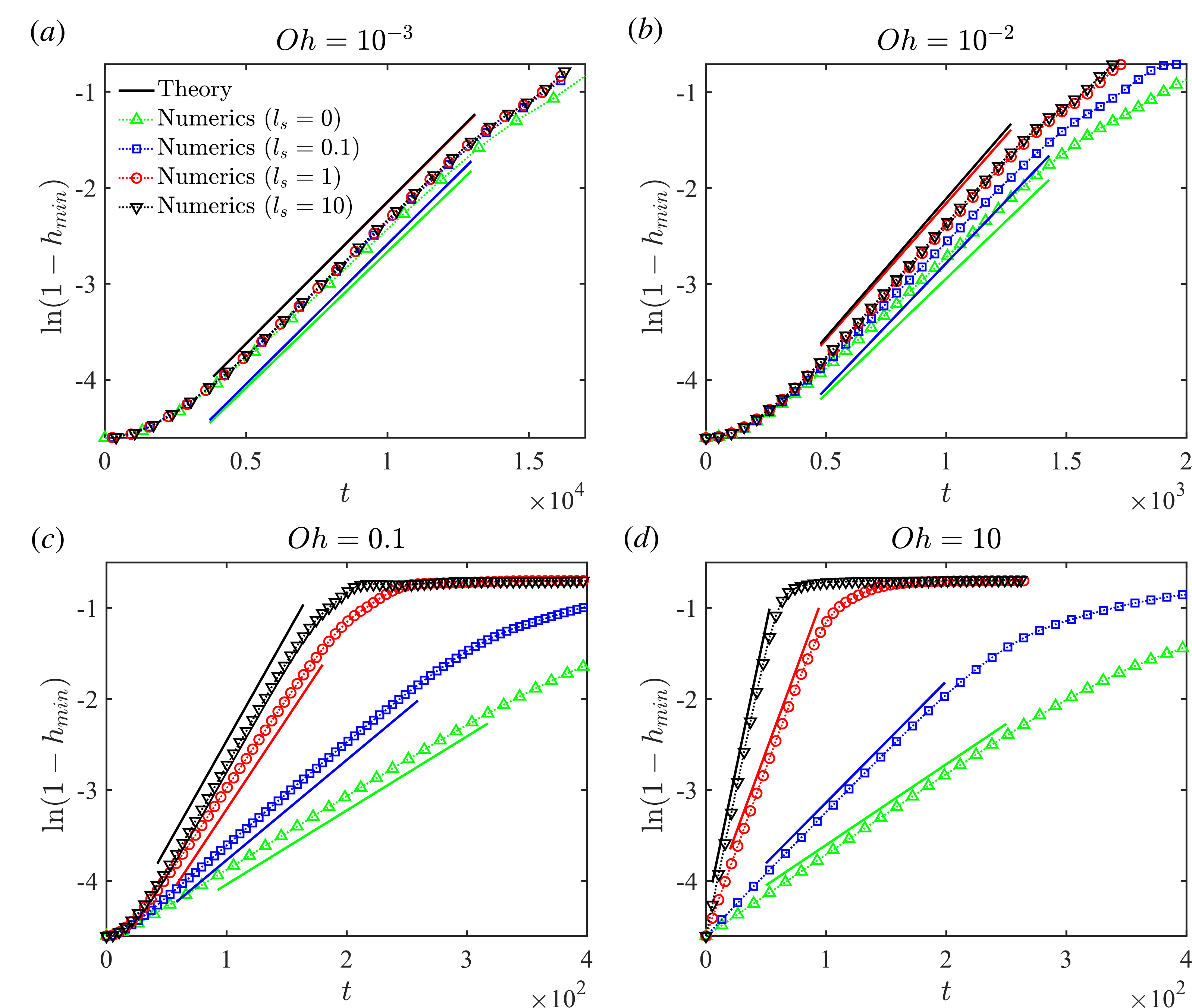}
	\caption{Linear time evolution of the minimum radii of films $h_{min}(t)$ with four $Oh$ values: (\textit{a}) $Oh=10^{-3}$, (\textit{b}) $Oh=10^{-2}$, (\textit{c}) $Oh=0.1$, (\textit{d}) $Oh=10$. The numerical solutions (dotted lines with symbols) are compared to the predictions of the NS dispersion relation (solid lines) for three four lengths: $l_\mathrm{s} = 0$ (green), $0.1$ (blue), $1$ (red) and $10$ (black). 
The theoretical values of $\omega$ are: (\textit{a})  $2.81 \times 10^{-4}$ (green), $2.91 \times 10^{-4}$ (blue), $2.94 \times 10^{-4}$ (red), $2.95 \times 10^{-4}$ (black); 
(\textit{b}) $2.41 \times 10^{-3}$ (green), $2.61 \times 10^{-3}$ (blue), $2.84 \times 10^{-3}$ (red), $2.89 \times 10^{-3}$ (black);
(\textit{c}) $8.12 \times 10^{-3}$ (green), $1.10 \times 10^{-2}$ (blue), $1.99 \times 10^{-2}$ (red), $2.34 \times 10^{-2}$ (black);
(\textit{d}) $8.90 \times 10^{-3}$ (green), $1.35 \times 10^{-2}$ (blue), $3.62 \times 10^{-2}$ (red), $6.32 \times 10^{-2}$ (black).
}
\label{fig_linear_growth}	
\end{figure}

Figure\,\ref{fig_linear_growth} illustrates the growth rates of perturbations in the twenty simulated cases.
Based on the instability analysis (\S\,\ref{subsec_dis_relation}) that employs the expression $h(z,t) = 1+\hat{h}e^{ikz+\omega t}$, the initial perturbation, modelled by a cosine function with a fixed wavenumber $k=2 \pi/L$, experiences exponential growth.
So $\mathrm{ln}(1-h_{min})$ is utilised as the $y$-coordinate in figure\,\ref{fig_linear_growth} to present the linear growth of perturbations.
The numerical results closely align with the theoretical predictions, demonstrating the effectiveness of the NS dispersion relation (\ref{eq_full_dispersion_relation}) in describing the inertial effects on the instability of films on slippery fibres.
Furthermore, it becomes evident that deviations due to slip become more pronounced as inertial effects diminish, thus providing quantitative confirmation of previous findings.

\begin{figure}
\centering
\includegraphics[width=1.0\textwidth]{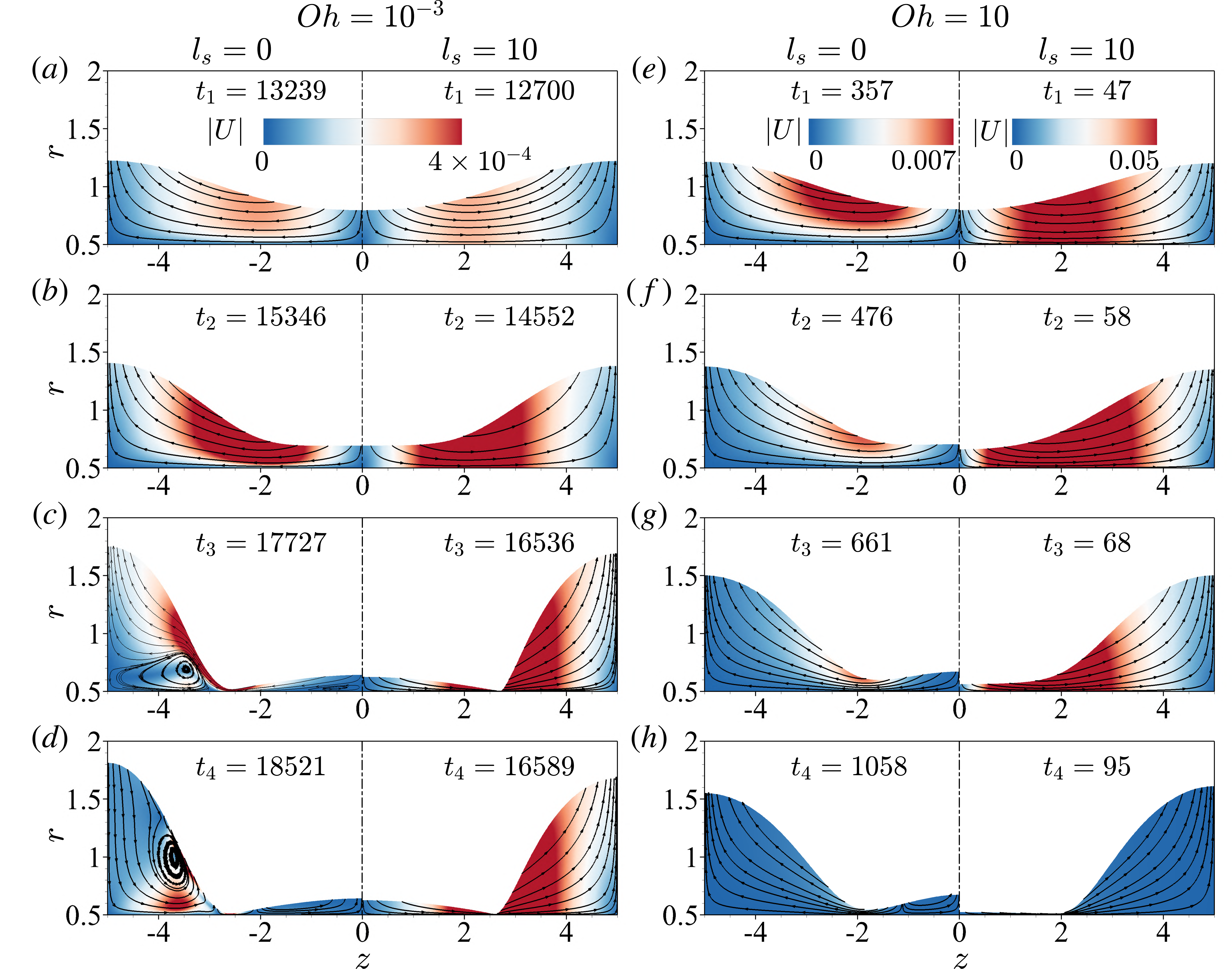}
	\caption{Evolutions of perturbation growth at the nonlinear stage with two $Oh$ values: (a-d) $Oh=10^{-3}$, (e-h) $Oh=10$. 
	The contours represent the velocity magnitude $\left| U\right| = \sqrt{u^2+w^2}$ .
The left panels in each figure illustrate the contours and streamlines of no-slip cases.The right panels show the results of the slip cases ($l_s=10$).
	}
\label{fig_flow_field_nonlinear}	
\end{figure}

In addition to examining the evolution of perturbation growth at the linear stage, we also delve into the nonlinear dynamics, illustrated in figure\,\ref{fig_flow_field_nonlinear}. 
Due to the dramatic changes in the interface profile at the nonlinear stage, the initial dense mesh experiences significant deformation, resulting in poor grid quality and potential numerical errors.
To address this challenge, we implement an approach of remeshing, regenerating  the mesh (reducing nodes) in the vicinity near the point of $h_{min}$ when $h_{min}<0.55$.
The simulations encompass both inertia-dominated and viscosity-dominated cases on the no-slip (left panel) and slip (right panel) boundary conditions.
Notably, the nonlinear evolution reveals substantial distinctions from the dynamics at the linear stage.
The interface shapes are found to deviate from their initial cosine, forming plateau structures at their lowest points. 
Ultimately, satellite droplets emerge between the two main drops.
Additionally, fluid structure within the film no longer exhibits `similarity' at different time instants owing to the drastic changes in interface profiles. 
In scenarios dominated by inertia, although slip has a minimal effect on the interface shape, it substantially alters the flow structure. 
Figures\,\ref{fig_flow_field_nonlinear}\,(\textit{c,d}) illustrate the evolution of vortical structures within the liquid film on a no-slip wall, resulting in significant oscillations of the interface before rupture. 
Conversely, no vortices appear within the film on the slippery fibre due to the weak shear forces acting on the fluid near the surface.
In viscosity-dominated scenarios, slip not only accelerates perturbation growth significantly, as observed at the linear stage, but also affects the interface profiles near rupture. This alteration results in the formation of filaments, rather than satellite droplets, between the two main drops. 
One plausible explanation is that the dominant perturbation wavelength of the instability in the slip case ($\lambda_{max}=15.32$) is notably larger than that in the no-slip case ($\lambda_{max}=9.38$), resulting in the flatter profile in figure\,\ref{fig_flow_field_nonlinear}\,(\textit{h}).
The observation also suggests that the volume of satellite droplets is influenced by both inertia and slip.

\begin{figure}
\centering
\includegraphics[width=1.0\textwidth]{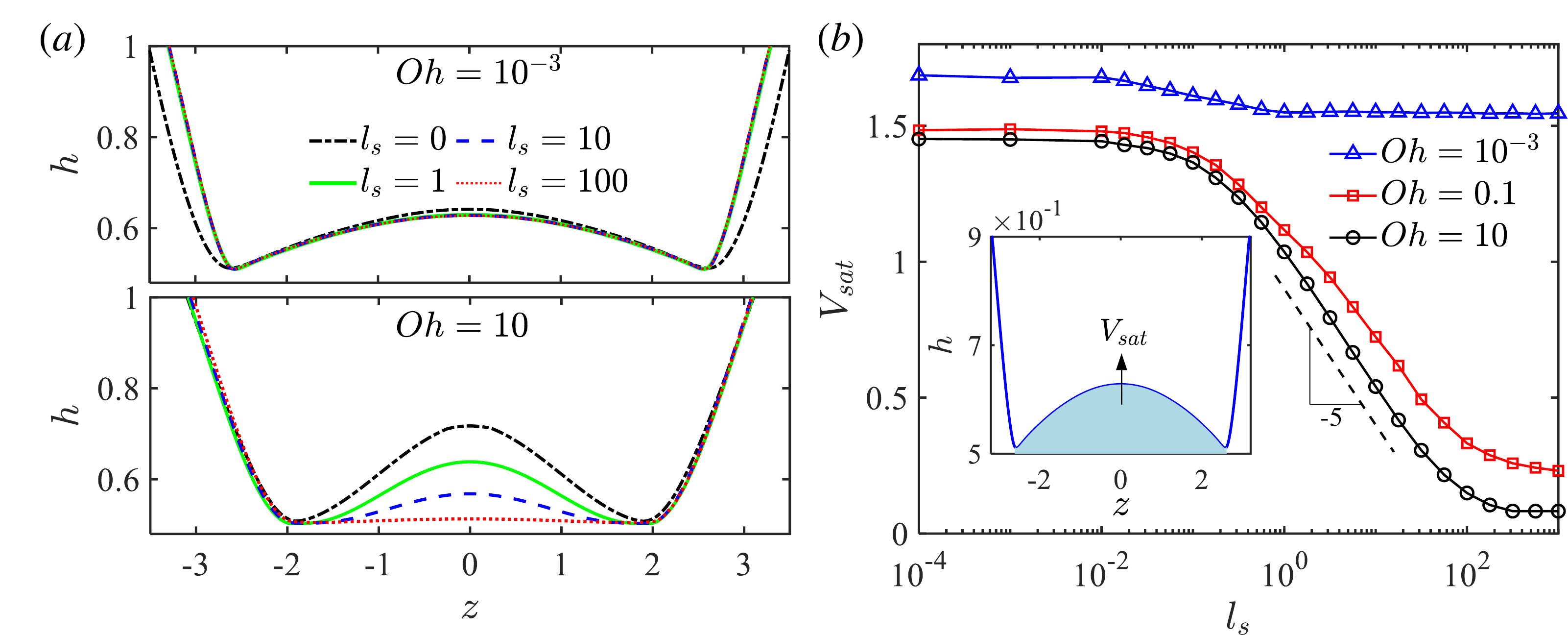}
	\caption{(\textit{a}) The interface profiles of satellite droplets on different boundary conditions with: $l_s=0$ (black dash-dotted lines), $1$ (green solid lines), $10$ (blue dashed lines) and $100$ (red dotted lines).  The upper and lower panel shows that the results with $Oh=10^{-3}$ and $Oh=10$, respectively. (\textit{b}) Variations of the satellite droplets with slip lengths. The inset provides a schematic of the volume of satellite droplets. }
\label{fig_vsat}	
\end{figure}

Figure\,\ref{fig_vsat} presents variations of satellite droplets, extracted from more than seventy simulations with varying values of $Oh$ and $l_s$.
In figure\,\ref{fig_vsat}\,(\textit{a}), we depict the interface profiles of satellite droplets. These profiles clearly demonstrate that in viscosity-dominated scenarios, slip significantly reduces the volumes of satellite droplets. However, in the case of inertia-dominated scenarios where $l_s>1$, slip has no discernible effect on the droplet volumes. 
Note that these profiles were obtained when $h_{min} \leq 0.01\,h_0$.
Furthermore, the volume of the satellite droplets is quantified  by calculating the area between the two lowest points of the profiles, as shown in figure\,\ref{fig_vsat}\,(\textit{b}).
It is evident that in viscous cases, the volume of satellite droplets decreases as $l_s$ increases, and higher viscosity (larger $Oh$) leads to a more rapid rate of decrease. 
When viscosity dominates the fluids ($Oh>=10$), the relationship between volume ($V_{sat}$) and slip length ($l_s$) approximately follows $V_{sat} \sim l_s^{-5}$.

\section{Conclusions}
In this study, a theoretical model is developed based on the linear instability analysis of the axisymmetric NS equations to investigate the influence of inertia and slip on dynamics of liquid films on fibres. 
The model is verified theoretically by the limiting cases for jet flows and film flows without inertia.
The resulting dispersion relation (\ref{eq_full_dispersion_relation})  reveals some intriguing insights. Firstly, it shows that slip has a relatively minor impact on the instability of flows dominated by inertia, whereas it significantly accelerates the growth of perturbations in viscous film flows. 
Moreover, the influence of slip appears to be contingent on the thickness of the liquid film, with thinner films exhibiting more pronounced slip effects.
We also extract the dominant perturbation modes, denoted as $k_{max}$, from the dispersion relation. In cases with no-slip boundary conditions, $k_{max}$ remains largely unaffected by inertia. Remarkably, for thin films characterized by $\alpha=0.8$, $k_{max}$ maintains a nearly constant value, closely aligning with the predictions of the no-slip lubrication model (\ref{eq_LE_noslip}), even as $Oh$ varies.
Conversely, when slip is introduced, $k_{max}$ exhibits a noticeable decrease as inertia decreases. In the limiting case where $l_s \rightarrow \infty$, the giant-slip lubrication model \eqref{eq_giant_LE} offers an approximate prediction for $k_{max}$.

To substantiate our theoretical findings, direct numerical simulations of the NS equations are conducted via using two distinct fluid configurations: (i) a long film with random initial perturbations to investigate the dominant modes of perturbations and (ii) a short film with a fixed wavelength to examine the evolution of perturbation growth.
The velocity fields extracted from these simulations yield valuable insights. Notably, the parabolic field of the velocity profiles in inertia-dominated cases are observed to be significantly smaller than those in cases governed by viscosity. Given that slip primarily influences non-uniform velocity profiles, this observation provides an  explanation for why the impact of slip is dampened by inertia.
Furthermore, we delve into the realm of nonlinear dynamics in perturbation growth. We found that nonlinear dynamics, in contrast to the linear stage, leads to the generation of intricate vortical structures near no-slip surfaces in inertia-dominated cases. Interestingly, slip was found to mitigate the impact of these shear stresses, resulting in smoother flows without the presence of vortices.
In the context of viscous cases, slip was identified to reduce the volume of satellite droplets ($V_{sat}$) at the final stages before rupture.
When viscosity dominates the fluid dynamics, this reduction in volume followed an approximate power law relationship, specifically $V_{sat} \sim l_s^{-5}$.

Given that experimental evidence has already confirmed the existence of wall slip in films on fibres, as demonstrated in previous studies \citep{haefner2015influence,ji2019dynamics}, and that slip length can be directly quantified through methods such as those described in \citep{huang2006direct,maali2012measurement,maali2016slip}, it is our hope that the wall slip can be controlled
experimentally to validate our predictions, especially regarding the impact of inertia on wavelengths (drop size) for various liquids. For instance, experiments involving water can be conducted to examine inertia-dominated scenarios, while experiments with silicone oil can be carried out to investigate viscosity-dominated cases.
There are numerous potential extensions to this framework. One avenue is to incorporate the influence of other physical factors, such as electric fields \citep{ding2014dynamics} and intermolecular forces \citep{ji2019dynamics,tomo2022observation}, both of which are known to affect critical wavenumbers ($k_{crit}$) and dominant modes ($k_{max}$) in the instability. Additionally, exploring related flow configurations, such as a liquid film flowing down a fibre driven by a body force, as discussed in \citep{liu2021coating}, presents intriguing opportunities. An open problem in this context is to comprehend the impact of inertial effects on the dynamics of traveling waves in various flow regimes.

%\section*{Acknowledgements}~~~~~~~~~~~~~~~~~~~~~~~~~~~~~~~
\backsection[Acknowledgements]{
Useful discussions with Dr. Yixin Zhang, Dr. Zhaodong Ding and Mr. Jingkun Xiong are gratefully acknowledged.}

%funding
\backsection[Funding]{This work was supported by the National Natural Science Foundation of China (grant no. 12202437, 12027801, 12388101, 12272372), the Youth Innovation Promotion Association CAS (grant no. 2018491, 2023477), the Fundamental Research Funds for the Central Universities, the China Postdoctoral  Science Foundation (grant no. 2022M723044) and the China International Postdoctoral Exchange Program Fellowship (grant no. YJ20210177) } 
%\section*{Declaration of interests}~~~~~~~~~~~~~~~~~~~~~~~
\backsection[Declaration of interests]{
The authors report no conflict of interest.}

\backsection[Author ORCIDs]{\\
Chengxi Zhao https://orcid.org/0000-0002-3041-0882\\
Ran Qiao https://orcid.org/0000-0002-4445-1255\\
Kai Mu https://orcid.org/0000-0002-4743-2332\\
Ting Si https://orcid.org/0000-0001-9071-8646 \\
Xisheng Luo  https://orcid.org/0000-0002-4303-8290
}

\appendix
\section{Derivation for the giant-slip lubrication model \label{app_deri_LE}}
In this appendix, we follow the approach employed by \citet{munch2005lubrication} to derive the giant-slip lubrication model.
To get this lubrication equation from the axisymmetric Navier-Stokes equations, we need to establish the leading order terms by their asymptotic expansion in $\varepsilon$, for which we use the rescaling shown below: 
\begin{align}
\tilde{x}= \tilde{\lambda} x,\,\, \tilde{r}=\varepsilon \tilde{\lambda} r,\,\, 
\tilde{w} = \tilde{U}w,\,\,
\tilde{u}=\varepsilon  \tilde{U} u,\,\,
\tilde{t} = \frac{\tilde{\lambda}}{\tilde{U}} t,\,\,
\tilde{p} = \frac{\mu U }{ \lambda } p\,. 
\end{align}
Here, $ \tilde{\lambda} = h_0 / \varepsilon  $. 
Substituting all these scalings into the dimensional NS equations yields,
 \begin{equation}
\frac{\partial w}{\partial z} + \frac{1}{r} \frac{\partial(u r) }{\partial r}  = 0\,, \\
 \end{equation}
    \begin{equation}
    \varepsilon^2 Re\left( \frac{\partial u}{\partial t}+ w \frac{\partial u}{\partial z} + u \frac{\partial u}{\partial r} \right)
    =  - \frac{\partial p }{\partial r}+ \varepsilon^2 \frac{\partial^2 u}{\partial z^2} 
    +\frac{\partial}{\partial r} \left[\frac{1}{r} \frac{\partial (ur)}{\partial r}\right]  \,. 
  \end{equation}
  \begin{equation}
  \label{eq_app_momentum}
\varepsilon^2 Re\left( \frac{\partial w}{\partial t}+ w \frac{\partial w}{\partial z} + u \frac{\partial w}{\partial r} \right)= -\varepsilon^2 \frac{\partial p }{\partial z}+ \varepsilon^2 \frac{\partial^2 w}{\partial z^2} 
  + \frac{1}{ r} \frac{\partial}{\partial r}\left(r \frac{\partial w}{\partial r} \right)  \,,
  \end{equation}
The equations at the liquid-gas interface ($r=h$) are scaled as
  \begin{equation}
     \frac{\partial h}{\partial t} + w \frac{\partial h}{\partial z} -u= 0 \,,
\end{equation}     
  \begin{equation}
  \label{eq_scaling_5}
  \begin{split}
  p - \frac{2}{1+ \varepsilon^2 \left(\partial_z h \right)^2} 
  \left[ \frac{\partial u}{\partial r} - \frac{\partial h}{\partial z} \left(  \frac{\partial w}{\partial r} 
  + \varepsilon^2 \frac{\partial u}{\partial z}\right) + \varepsilon^2 \left(\frac{\partial h}{\partial z} \right)^2 \frac{\partial w}{\partial z} \right]  \\
 =\frac{Re}{We} \left[ \frac{1}{\varepsilon^2 h \sqrt{1+ \varepsilon^2 \left(\partial_z h \right)^2}}  - \frac{\partial_z^2 h}{\left(1+\varepsilon^2 \left(\partial_z h \right)^2\right) ^\frac{3}{2}} \right]  \,,
  \end{split}
  \end{equation} 
  \begin{equation}
  \label{eq_scaling_6}
  2 \varepsilon^2 \frac{\partial h}{\partial z} \left( \frac{\partial u}{\partial r} - \frac{\partial w}{\partial z} \right) 
  + \left[ 1- \varepsilon^2 \left(\frac{\partial h}{\partial z} \right)^2 \right]  \left( \frac{\partial w}{\partial r} + \varepsilon^2 \frac{\partial u}{\partial z} \right) = 0 \,.
  \end{equation}
For the boundary conditions at the liquid-solid interface ($r=\alpha$) the scaled form are
\begin{align}
\label{eq_scaling_7}
 & w = \frac{ l_{s}}{ \varepsilon^2 } \frac{\partial w}{\partial r}\,, \\ 
 &u=0 \,.
\end{align}

The rescaled equations can be approximately solved by the perturbation expansion, expressed as:
\begin{equation}
\left( w,u,p,h\right) = \left( w_0,u_0,p_0,h_0\right) + \varepsilon^2   \left( w_1, u_1, p_1, h_1\right)+ ...
\end{equation} 
After eliminating all the high order terms of $\varepsilon$ and only keeping the terms of the leading order, we obtain
 \begin{equation}
  \label{eq_rescale_mass}
 \partial_z w_0+ \partial_r(u_0 r)/r = 0 \,, 
 \end{equation}
    \begin{equation}
    \label{eq_rescale_r_momen}
  \partial_r p_0 = \partial_r \left[ \partial_r \left(u_0 r \right) / r\right]    \,,
  \end{equation}
  \begin{equation}
    \label{eq_rescale_x_momen}
\partial_r \left(r \partial_r w_0 \right) =0\,, 
  \end{equation}
  \begin{equation}
      \label{eq_rescale_kine}
     \partial_{t} h_0  + w_0\, \partial_z h_0  - u_0 = 0 \,,
\end{equation}     
\begin{equation}
\label{eq_rescale_nor}
 p_0 -2\left( \partial_r u_0  - \partial_z h_0 \,\partial_r w_0 \right)   = \frac{Re}{ We } \left( 1/h_0  - \partial_z^2 h_0 \right),  \quad (r=h) \,,
 \end{equation}
\begin{equation}
\label{eq_rescale_tange}
 \partial_r w_0 = 0, \quad (r=h) \,,
\end{equation}
\begin{equation}
\label{eq_rescale_slip}
 \partial_r w_0 = 0, \quad (r=\alpha) \,,
\end{equation}
\begin{equation}
\label{eq_rescale_no_pen}
 u_0 = 0,  \quad (r=\alpha) \,.
\end{equation}
According to the equations\,(\ref{eq_rescale_x_momen}), (\ref{eq_rescale_slip}) and (\ref{eq_rescale_no_pen}), $w$ is independent of the $r$, i.e. $w_0 = w_0(z,t)$.
So equation\,(\ref{eq_rescale_mass}) is rearranged as
\begin{equation}
u_0 = - r \partial_z w_0/2 \,.
\label{eq_rescaled_mass_rearranged}
\end{equation}
Substituting (\ref{eq_rescaled_mass_rearranged}) into (\ref{eq_rescale_kine}) and (\ref{eq_rescale_nor}) yields
\begin{align}
\label{eq_LE_first_order_h}
& \partial_{t} h_0  + w_0\, \partial_z h_0  + h_0 \partial_z w_0/2= 0\,,  \\
\label{eq_LE_first_order_p}
&  p_0+\partial_z w_0 = \frac{Re}{ We } \left(  1/h_0 - \partial_z^2 h_0 \right) \,,
\end{align}

For the terms of the next order, equation\,(\ref{eq_app_momentum}) becomes
\begin{equation}
\label{eq_LE_next_order_ori}
 Re\left( \frac{\partial w_0}{\partial t}+ w_0 \frac{\partial w_0}{\partial z}  \right)=- \frac{\partial p_0 }{\partial z}+ \frac{\partial^2 w_0}{\partial z^2} 
  + \frac{1}{ r} \frac{\partial}{\partial r}\left(r \frac{\partial w_1}{\partial r} \right)  \,,
\end{equation}
and equations\,(\ref{eq_scaling_6}) and (\ref{eq_scaling_7}) are applied for the boundary conditions
\begin{align}
\label{eq_LE_next_order_BC1}
  -3 \frac{\partial h_0}{\partial z} \frac{\partial w_0}{\partial z} 
  + \frac{\partial w_1}{\partial r}-\frac{r}{2}\frac{\partial^2 w_0}{\partial z^2} &= 0\,, \quad (r=h) \,, \\
  \label{eq_LE_next_order_BC2}
  w_0 - l_{s} \frac{\partial w_1}{\partial r} &= 0 \,, \quad (r=\alpha) \,,
\end{align}
Integrating (\ref{eq_LE_next_order_ori}) from $\alpha$ to $h$ and using boundary conditions (\ref{eq_LE_next_order_BC1}) and (\ref{eq_LE_next_order_BC2}) leads to
\begin{equation}
\label{eq_LE_next_order_int}
 Re\left( \frac{\partial w_0}{\partial t}+ w_0 \frac{\partial w_0}{\partial z}  \right) \int_\alpha^h r dr= \left( -\frac{\partial p_0 }{\partial z}+ \frac{\partial^2 w_0}{\partial z^2} \right) \int_\alpha^h r dr
  +r \frac{\partial w_1}{\partial r}  \bigg|_{r=\alpha}^{r=h} \,,
\end{equation}
Rearranging (\ref{eq_LE_next_order_int}) yields
\begin{equation}
\label{eq_LE_next_order_result}
 Re\left( \frac{\partial w_0}{\partial t}+ w_0 \frac{\partial w_0}{\partial z}  \right)=  -\frac{\partial p_0 }{\partial z}+ \frac{\partial^2 w_0}{\partial z^2} 
 + \frac{h_0}{h_0^2-\alpha^2}\left( 6 \frac{\partial h_0}{\partial z} \frac{\partial w_0}{\partial z} + h_0 \frac{\partial^2 w_0}{\partial z^2}\right) - \frac{2 \alpha}{h_0^2-\alpha^2} \frac{w_0}{l_{s}}\,.
\end{equation}
%\newpage
%\bibliographystyle{unsrt}
%\bibliography{thin_film_equation}

Combining (\ref{eq_LE_first_order_h}), (\ref{eq_LE_first_order_p}) and (\ref{eq_LE_next_order_result}) gives the final lubrication model  for the film on a fibre with a giant slip length, written as
\begin{subequations}\label{eq_giant_LE2}
	\begin{empheq}[left={\empheqlbrace}]{alignat=1}
		\frac{\partial h^2}{\partial t} + \frac{\partial \left(h^2 w \right)}{\partial z} = & 0\,, \\
		\frac{\partial w}{\partial t}+ w \frac{\partial w}{\partial z}  = & -\frac{1}{ We} \frac{\partial}{\partial z}\left(\frac{1}{h}-\frac{\partial^2 h}{\partial z^2} \right)+\frac{1}{ Re}  \frac{3}{h^2-\alpha^2} \frac{\partial \left( h^2 \partial_z w\right)}{\partial z} \notag\\
		&- \frac{1}{ Re} \frac{2 \alpha^2}{h^2-\alpha^2}  \left(\frac{\partial^2 w}{\partial z^2}-\frac{w}{ \alpha\, l_{s}} \right)\,. \label{eq_giant_slip_mom2}
	\end{empheq}
\end{subequations}
If $U = \gamma / \mu$, $Re = We = Oh^{-2}$, the momentum equation (\ref{eq_giant_slip_mom2}) becomes
\begin{equation}
 \frac{\partial w}{\partial t}+ w \frac{\partial w}{\partial z}  =  - Oh^2 \frac{\partial}{\partial z}\left(\frac{1}{h}-\frac{\partial^2 h}{\partial z^2} \right) 
+ Oh^2 \frac{3}{h^2-\alpha^2} \frac{\partial \left( h^2 \partial_z w\right)}{\partial z} - Oh^2 \frac{2 \alpha^2}{h^2-\alpha^2}  \left(\frac{\partial^2 w}{\partial z^2}-\frac{w}{ \alpha\, l_\mathrm{s}} \right)\,.
\end{equation}

It is worth noting that $h^2-\alpha^2$ introduces singularity as $h \rightarrow \alpha$. A similar singularity is also observed in other lubrication equations describing free-surface flows with significant inertial effects, such as liquid jets \citep{eggers1994drop}, liquid sheets \citep{erneux1993nonlinear}, and planar films on ultra-slip walls \citep{munch2005lubrication}. However, this singularity disappears in lubrication models that describe inertialess flows of bounded thin films \citep{kang2017marangoni,RN99}.
The absence is primarily attributed to the distinct scaling used in the lubrication approximation.
	
\section{Linear instability analysis for lubrication equations}
In this appendix, the instability analysis is performed for the lubrication equations \eqref{eq_giant_LE} and \eqref{eq_LE_noslip} using the normal mode method.

For the giant-slip model \eqref{eq_giant_LE}, substituting $h(z,t) = 1 + \hat{h} e^{\omega t + i k z}$ and $w(z,t) = \hat{w} e^{\omega t + i k z}$ into the linearised lubrication equation gives
  \begin{align}
& \omega \hat{h} = - ik \hat{w}/2\,, \\
\label{eq_LSA_momentum}
& Oh^{-2}\, \omega \hat{w}=  ik(1-k^2) \hat{h}
- \frac{3 k^2 \hat{w}}{1-\alpha^2}+ \frac{2 \alpha^2}{1-\alpha^2}  \left(k^2+ \frac{1}{\alpha\, l_\mathrm{s}} \right) \hat{w}\,.
\end{align}
Eliminating the $\hat{h}$ in (\ref{eq_LSA_momentum}) yields the dispersion relation
\begin{equation}
\label{eq_dispersion_giant_slip}
\omega^2 + Oh^2 \left[ \frac{3-2\alpha^2}{1-\alpha^2}k^2 - \frac{2 \alpha}{(1-\alpha^2) l_\mathrm{s}}\right] \omega + Oh^2 \frac{k^2 (k^2-1)}{2}=0
\end{equation}
With a similar approach, we can have the dispersion relation from the no-slip lubrication model (\ref{eq_LE_noslip}), namely
\begin{equation}
\label{eq_dispersion_LE}
\omega = (k^2-1)  k^2 (3+ \alpha^4 - 4 \alpha^2 + 4\, \mathrm{ln}\,\alpha )/16.
\end{equation}

\section{Auxiliary functions for the dispersion relations \label{app_func}}
In this appendix, we present the definitions of the auxiliary functions for the dispersion relations \eqref{eq_Stokes_DR1} and \eqref{eq_dispersion_stokes}.

The $G_{ij}$ in \eqref{eq_Stokes_DR1} are
\begin{empheq}[left={\empheqlbrace}]{alignat=1}
G_{21} & = k^2 \mathrm{I}_0(k \alpha)-l_{s} k^3 \mathrm{I}_1(k\alpha),\nonumber 
\\
G_{22} & = k \mathrm{I}_0(k\alpha) + k^2 \alpha \mathrm{I}_0'(k\alpha)-2 l_s k^2 \mathrm{I}_1(k\alpha) - l_s k^3 \alpha \mathrm{I}_1'(k\alpha), \nonumber
\\
G_{23} &= k^2 \mathrm{K}_0(k \alpha)+l_{s} k^3 \mathrm{K}_1(k \alpha) ,\nonumber
\\
G_{24} & = - k \mathrm{K}_0(k\alpha) - k^2 \alpha \mathrm{K}_0'(k\alpha)-2 l_s k^2 \mathrm{K}_1(k\alpha) - l_s k^3 \alpha \mathrm{K}_1'(k\alpha), \nonumber 
\\
G_{41} &= 2k^2 \mathrm{I}'_1(k) \omega + (k^2-1) k \mathrm{I}_1(k), \nonumber
\\
G_{42} & = 2\left[ k^2 \mathrm{I}_0'(k)-  k \mathrm{I}_1'(k) \right] \omega+ (k^2-1)k \mathrm{I}_1'(k)\,, \nonumber 
\\
G_{43} &=  - 2k^2 \mathrm{K}'_1(k) \omega - (k^2-1) k \mathrm{K}_1(k) \,, \nonumber 
\\
G_{44} & = -  2\left[ k^2 \mathrm{K}_0'(k)+  k \mathrm{K}_1'(k) \right] \omega+ (k^2-1)k \mathrm{K}_1'(k)\,. \nonumber 
\end{empheq}

For \eqref{eq_dispersion_stokes}, we have
\begin{align}
	\Delta_1& = 
	\begin{array}{|ccc|}
		\alpha \mathrm{I}_0(k \alpha) 
		& \mathrm{K}_1(k\alpha) 
		& \alpha \mathrm{K}_0(k \alpha)  \\ 
		H_{22}
		&H_{23}
		& H_{24}\\
		k \mathrm{I}_0(k) +\mathrm{I}_1(k)
		& k \mathrm{K}_1(k) 
		&  k \mathrm{K}_0(k)-\mathrm{K}_1(k) \\
	\end{array}\,, \nonumber \\ \nonumber
	\Delta_2 &= 
	\begin{array}{|ccc|}
		\mathrm{I}_1(k\alpha) 
		& \mathrm{K}_1(k\alpha) 
		& \alpha \mathrm{K}_0(k \alpha)  \\ 
		H_{21}
		&H_{23}
		& H_{24}\\
		k \mathrm{I}_1(k ) 
		& k \mathrm{K}_1(k) 
		&  k \mathrm{K}_0(k)-\mathrm{K}_1(k) \\
	\end{array}\,, \\ \nonumber
	\Delta_3 &= 
	\begin{array}{|ccc|}
		\mathrm{I}_1(k\alpha) 
		&  \alpha \mathrm{I}_0(k \alpha) 
		& \alpha \mathrm{K}_0(k \alpha)  \\ 
		H_{21}
		& H_{22}
		& H_{24}\\
		k \mathrm{I}_1(k ) 
		& k \mathrm{I}_0(k) +\mathrm{I}_1(k)
		&  k \mathrm{K}_0(k)-\mathrm{K}_1(k) \\
	\end{array}\,, \\ \nonumber
	\Delta_4 &= 
	\begin{array}{|ccc|}
		\mathrm{I}_1(k\alpha) 
		&  \alpha \mathrm{I}_0(k \alpha) 
		& \mathrm{K}_1(k\alpha)  \\ 
		H_{21}
		& H_{22}
		&H_{23}\\
		k \mathrm{I}_1(k ) 
		& k \mathrm{I}_0(k) +\mathrm{I}_1(k)
		& k \mathrm{K}_1(k)  \\
	\end{array}\,,  
\end{align}
where
\begin{empheq}[left={\empheqlbrace}]{alignat=1}
	H_{21} &= k \mathrm{I}_0(k \alpha)-l_{s} k^2 \mathrm{I}_1(k\alpha) \,, \nonumber
\\
H_{22} & =  (2- l_s k^2 \alpha ) \mathrm{I}_0(k\alpha) +(\alpha-2 l_s) k \mathrm{I}_1(k\alpha)\,, \nonumber
\\ 
H_{23} &=  -k \mathrm{K}_0(k \alpha)-l_{s} k^2 \mathrm{K}_1(k \alpha) \,, \nonumber
\\
H_{24} & =(2-l_s k^2 \alpha) \mathrm{K}_0(k\alpha) + (2 l_s-\alpha)k \mathrm{K}_1(k\alpha)\,. \nonumber 
\end{empheq}

\section{Comparisons of the interface profiles}\label{app_com_int}
\begin{figure}
	\centering
	\includegraphics[width=0.9\textwidth]{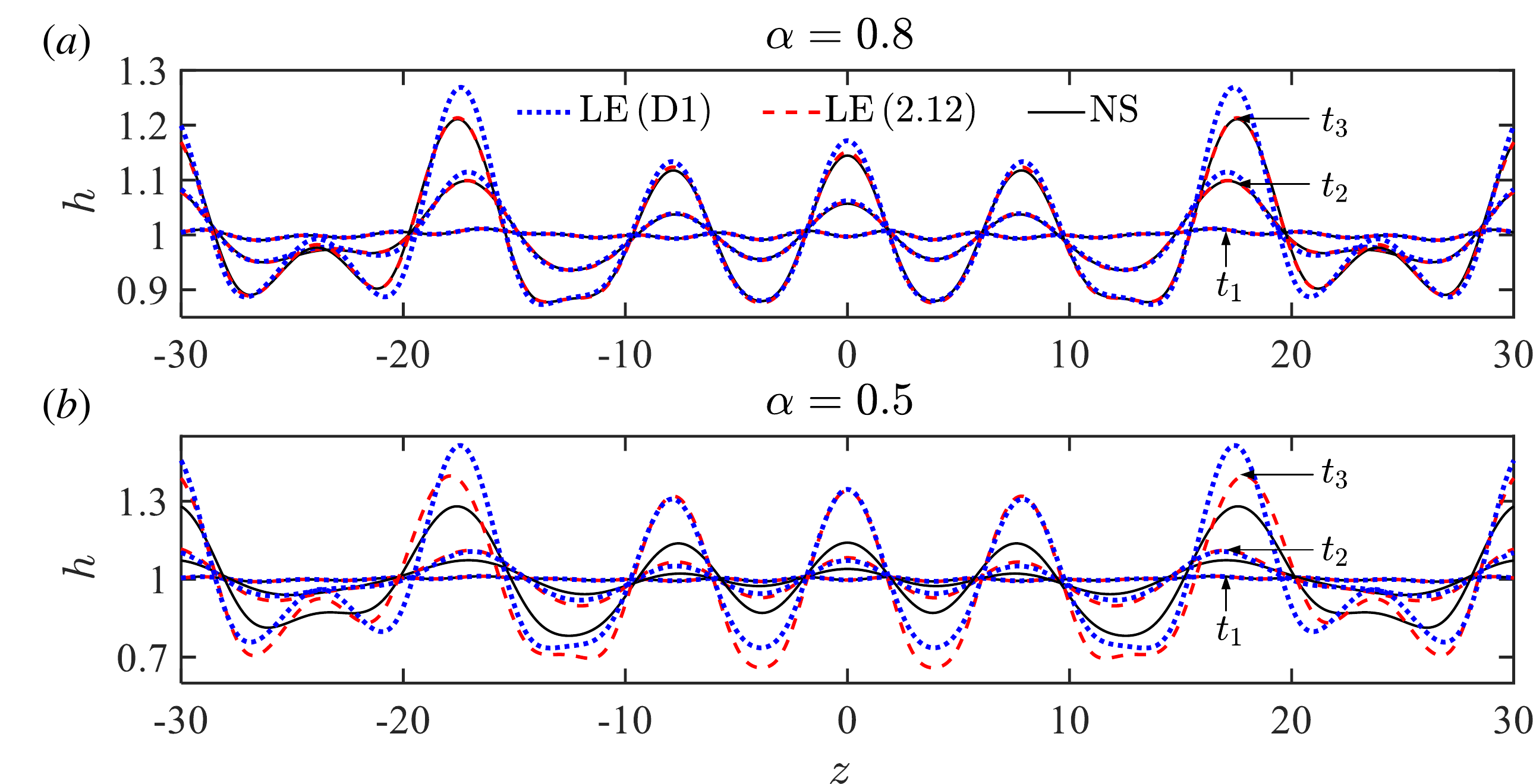}
	\caption{ Interface profiles of inertialess flows on two fibres of different radii at three time instants: (\textit{a}) $\alpha=0.8$, $t_1=0$, $t_2=5700$, $t_3=8400$; (\textit{b}) $\alpha=0.5$, $t_1=0$, $t_2=400$, $t_3=600$. 
	The blue dotted lines and red dashed lines represent the results predicted by \eqref{eq_LE_lister} and \eqref{eq_LE_noslip}, respectively. The solid lines are obtained from direct numerical simulations for the NS equations.}
	\label{fig_LE_NS}	
\end{figure}

In this appendix, we compare the interface profiles obtained through simulations for the NS equations with those calculated numerically from the lubrication equations (LEs).
Here, we focus on inertialess flows ($Oh \gg 1$) on no-slip fibres.
So we perform numerical investigations for the no-slip LE \eqref{eq_LE_noslip} introduced in \S\,\ref{sec_MM}.
Note that \citet{lister2006capillary} used a different LE \eqref{eq_LE_lister} to explore nonlinear dynamics of film interfaces, particularly those exhibiting characteristic collar-and-lobe structures, written as
\begin{equation}
	\label{eq_LE_lister}
\frac{\partial h}{\partial t} = -\frac{1}{ 3} \frac{\partial}{\partial z} \left[ (h-\alpha)^3 \left ( \frac{\partial h}{\partial z} + \frac{\partial^3 h}{\partial z^3}  \right)\right]\,.
\end{equation}
Here, the assumption $h-\alpha \ll \alpha$ simplifies the form of \eqref{eq_LE_lister} compared to \eqref{eq_LE_noslip}.
Both of these LEs, with periodic boundary conditions, are solved using a simple second-order finite-difference scheme in both time and space.
In the direct numerical simulations of the NS equations, we set $Oh=10$ and $l_s=0$ to enable a straightforward comparison with the numerical solutions of the LEs.
Figure\,\ref{fig_LE_NS}\,(\textit{a}) illustrates excellent agreement between the numerical predictions of the NS and LEs, starting from the same initial interface profile, for a thin film ($\alpha=0.8$). However, deviations become more pronounced for a thicker film ($\alpha=0.5$), consistent with previous theoretical results from instability analysis \citep{zhao2023slip}.

\bibliographystyle{jfm}
\bibliography{jfm}

\end{document}